\documentclass{birkjour_t2}
\usepackage{color}

 \theoremstyle{definition}
 
 \theoremstyle{remark}

\usepackage{subfigure}

\begin{document}

\title{Rogue waves and lumps on the non-zero background in the $\mathcal{PT}$-symmetric nonlocal Maccari system}

\author{Yulei Cao$^1$}

\address{$^1$ School of Mathematical Sciences, USTC, Hefei, Anhui 230026, P.\ R.\ China \\}

\email{caoyulei@mail.ustc.edu.cn}

\thanks{$^*$ Corresponding author: hejingsong@szu.edu.cn; jshe@ustc.edu.cn}
\author{Yi Cheng$^{1}$}
\address{$^1$ School of Mathematical Sciences, USTC, Hefei, Anhui 230026, P.\ R.\ China\\}
\email{chengy@ustc.edu.cn}

\author{Boris A. Malomed$^{2,3}$}
\address{$^2$ Department of Physical Electronics, School of Electrical Engineering, Faculty of Engineering, and Center for Light-Matter Interaction, Tel Aviv University, Tel Aviv 69978, Israel\\
$^3$ Instituto de Alta Investigaci$\acute{o}$n, Universidad de Tarapac$\acute{a}$, Casilla 7D, Arica, Chile\\}
\email{malomed@post.tau.ac.il}

\author{Jingsong He$^{4*}$}

\address{$^{4*}$ Institute for Advanced Study, Shenzhen University, Shenzhen,
Guangdong 518060, P.\ R.\ China \\}
\email{hejingsong@szu.edu.cn; jshe@ustc.edu.cn}

\begin{abstract}
In this paper, the $\mathcal{PT}$-symmetric version of
the Maccari system is introduced, which can be regarded as
a two-dimensional generalization of the defocusing nonlocal
nonlinear Schr$\rm\ddot{o}$dinger
equation. Various exact solutions of the nonlocal
Maccari system are obtained by means of the Hirota bilinear method,
long-wave limit, and Kadomtsev-Petviashvili (KP) hierarchy method.
Bilinear forms of the nonlocal Maccari system are derived for the first time. Simultaneously,
a new nonlocal Davey-Stewartson-type equation is derived. Solutions for breathers and
breathers on top of periodic line waves are obtained through the
bilinear form of the nonlocal Maccari system. Hyperbolic line rogue-wave
solutions and semi-rational ones,
composed of hyperbolic line rogue wave and periodic line
waves are also derived in the long-wave limit.
The semi-rational solutions exhibit a unique dynamical behavior. Additionally, general line soliton
solutions on constant background are generated by restricting
different tau-functions of the KP hierarchy, combined with the Hirota
bilinear method. These solutions exhibit elastic
collisions, some of which have never been reported before in nonlocal
systems. Additionally, the semi-rational solutions, namely
(i) fusion of line solitons and lumps into line solitons, and
(ii) fission of line solitons into lumps and line solitons,
are put forward in terms of the KP hierarchy. These novel semi-rational
solutions reduce to $2N$-lump solutions of the nonlocal Maccari
system with appropriate parameters. Finally, different characteristics
of exact solutions for the nonlocal Maccari system are summarized.
These new results enrich the structure of waves in nonlocal nonlinear systems,
and help to understand  new physical phenomena.
\end{abstract}

\maketitle


\vspace{-0.9cm}




\section{Introduction}
Nonlinear evolution equations serve as models for various complex physical
phenomena, therefore exact solutions of such equation have great significance\cite%
{ablow,hirota,waz}. In particular, rogue waves (RWs), as a kind of exact
solutions, have drawn much interest\cite{r1}-\cite{nrw2}. RW is
a special wave famous for its destructive force. Its amplitude can exceed
the background value by a factor of up to $3$\cite{pe}. RW was first derived
by Peregrine from the nonlinear Schr$\mathrm{\ddot{o}}$dinger equation\cite%
{pe}. Subsequently, a series of other soliton equations have been shown to
possess RW solutions\cite{guolj}-\cite{ma2}. Manifestations of
RWs have been identified in Bose-Einstein condensates\cite{be}, optical
systems\cite{os1,os2}, superfluids\cite{super}, and finances\cite{fi}.
Semi-rational solutions consisting of RWs, lumps, solitons and periodic line
waves also have attracted considerable attention\cite{s1}-\cite{semi2}. There
are several methods to obtain the semi-rational solutions of nonlinear evolution equations, including the Darboux transform and the bilinear method.

Nonlinear evolution equations may be naturally divided in two vast classes, \textit{viz}., local and
nonlocal ones. Recently, Ablowitz and Musslimani have introduced a nonlocal
reverse-space nonlinear Schr$\mathrm{\ddot{o}}$dinger (NLS) equation \cite%
{nls}
\begin{equation}
iu_{t}(x,t)-u_{xx}(x,t)-u^{2}(x,t)u^{\ast }(-x,t)=0.  \label{nNLS}
\end{equation}%
This equation generalizes the standard NLS equation by making it to satisfy
the parity-time ($\mathcal{PT}$) symmetry condition\cite{nls}. Ablowitz and Mussilimani have obtained
multi-soliton solutions of eq. \eqref{nNLS} by means of the inverse
scattering transform method\cite{nls}. Since then, the discrete versions\cite{dispt1}-\cite{dispt4} of eq. %
\eqref{nNLS} and its various solutions\cite{i1}-\cite{i6} have been studied in depth. Subsequently, several extended one-dimensional and
two-dimensional nonlocal nonlinear evolution equations have been introduced
\cite{mu1}-\cite{mu7}. An issue of obvious interest is to generate a
multidimensional integrable model with $\mathcal{PT}$-symmetry and explore
interaction dynamics in it, which is a motivation for the present work.

For nonlinear evolution equations with a dispersive linear part, wave
modulation induced by weak nonlinearity is a topic of general interest too,
with applications to plasma physics, nonlinear optics, hydrodynamics, etc. A
limited number of model equations play a crucially important role for this
purpose. They are produced by the reduction method applied to a family of
nonlinear evolution equations under suitable approximations
\cite{WE}-\cite{FC2}. Of special interest are integrable model equations. To
derive a (2+1)-dimensional integrable system with coordinates $\left(
x,y\right) $ and temporal variable $t$, Maccari\cite{mac} used a moving
reference frame,
\begin{equation}
\xi =\epsilon ^{p_{1}}(x-V_{1}t),\quad \eta =\epsilon
^{p_{2}}(y-V_{2}t),\quad \tau =\epsilon ^{q}t,  \label{xi}
\end{equation}%
where $p_{1},p_{2},q>0$ and $\mathbf{V(K)}\equiv \lbrack
V_{1}(K_{1},K_{2}),V_{2}(K_{1},K_{2})]$ is the vector of the group velocity
of the linearized equation, and $\epsilon$ is the expansion parameter, supposed to be sufficiently
small. In particular, starting from the celebrated
Kadomtsev-Petviashvili (KP) equation
\begin{equation}
U_{xt}(x,y,t)-3U_{xx}^{2}(x,y,t)+U_{xxxx}(x,y,t)+sU_{yy}(x,y,t)=0,
\label{KP}
\end{equation}%
$s$ is an arbitrary real constant, its linear dispersive part can be described in terms of Fourier modes, with
components of the group velocity
\begin{equation}
V_{1}(K_{1},K_{2})=-3K_{1}^{2}-s\frac{K_{2}^{2}}{K_{1}^{2}},\quad
V_{2}(K_{1},K_{2})=\frac{2sK_{2}}{K_{1}}.
\end{equation}%
Then, the following asymptotic Fourier expansion is introduced,
\begin{equation}
U(x,y,t)=\sum\limits_{n=-\infty }^{n=+\infty }\epsilon ^{\gamma _{n}}\psi
_{n}(\xi ,\eta ,\tau ;\epsilon )e^{in(K_{1}x+K_{2}y-\omega t)},  \label{sum}
\end{equation}%
where $\gamma _{n}=|n|$ and $\gamma _{0}=1+r$ are real constants, and $\psi
_{-n}(\xi ,\eta ,\tau ;\epsilon )=\psi _{n}^{\ast }(\xi ,\eta ,\tau
;\epsilon )$, because $U(x,y,t)$ is real. It is assumed that the limit of $\psi _{n}^{\ast }(\xi ,\eta ,\tau
;\epsilon )$'s, for $\epsilon\rightarrow 0$, exists and is finite.
The purpose of this
assumption is to derive an evolution equation for the modulation amplitude, $%
\psi _{1}(\xi ,\eta ,\tau ;\epsilon \rightarrow 0)\equiv \Psi (\xi ,\eta
,\tau )$.

If $p_{1}\neq p_{2}$ in eq. (\ref{xi}) for the proper balance of terms, we
get for $n=0$ (see eq. (\ref{sum}))
\begin{equation}
\Phi _{\tau }(\xi ,\eta ,\tau )+2\sqrt{3}K_{1}\Phi _{\eta }(\xi ,\eta ,\tau
)-6|\Psi (\xi ,\eta ,\tau )|_{\xi }^{2}=0,
\end{equation}%
and for $n=1$
\begin{equation}
i\Psi _{\tau }(\xi ,\eta ,\tau )-6K_{1}\Psi _{\xi \xi }(\xi ,\eta ,\tau
)+6K_{1}\Psi (\xi ,\eta ,\tau )\Phi (\xi ,\eta ,\tau )=0,
\end{equation}%
with $3K_{1}^{4}=K_{2}^{2},p_{1}=\frac{2}{3},p_{2}=\frac{4}{3},q=\frac{4}{3}%
,r=\frac{1}{3},\Phi _{\xi }(\xi ,\eta ,\tau )=\psi _{0}(\xi ,\eta ,\tau ).$
By means of obvious rescaling, the following KP-type system is produced\cite%
{mac,maccari}
\begin{eqnarray}
&&iu_{t}(x,y,t)+u_{xx}(x,y,t)+u(x,y,t)v(x,y,t)=0,  \notag \\
&&v_{t}(x,y,t)+v_{y}(x,y,t)+|u(x,y,t)|_{x}^{2}=0.  \label{kpb}
\end{eqnarray}%
Various exact solutions of local and nonlocal KP-type system\eqref{kpb} including solitons, breathers,
rational and semi-rational solutions have been reported \cite{ks1}-\cite{ks6}.

If $p_{1}=p_{2}$, for the proper balance of different terms, we get for $n=0$
\begin{equation}
V_{1}\Phi _{\xi \xi }+V_{2}\Phi _{\xi \eta }-s\Phi _{\eta \eta }+6|\Psi
|_{\xi \xi }^{2}=0,
\end{equation}%
and for $n=1$
\begin{equation}
iK_{1}\Psi _{\tau }-(6K_{1}+V_{1})\Psi _{\xi \xi }-V_{2}\Psi _{\xi \eta
}+s\Psi _{\eta \eta }+6k_{1}^{2}\Phi \Psi -6|\Psi |^{2}\Psi =0,
\end{equation}%
where $\Phi =\Phi (\xi ,\eta ,\tau ),\Psi =\Psi (\xi ,\eta ,\tau )$ ,$%
3K_{1}^{4}=K_{2}^{2},p_{1}=p_{2}=1,q=2,r=1,\Phi _{\xi }(\xi ,\eta ,\tau
)=\psi _{0}(\xi ,\eta ,\tau ).$ After a rescaling, the following \textit{%
Maccari system} is obtaioned\cite{mac}:%
\begin{eqnarray}
&&iu_{t}(x,y,t)+L_{1}u(x,y,t)+u(x,y,t)v(x,y,t)=0,  \notag \\
&&L_{2}v(x,y,t)-2L_{1}[u(x,y,t)u^{\ast }(x,y,t)]=0,  \label{maccari}
\end{eqnarray}%
where $v(x,y,t)$ is a real function and linear differential operators $L_{1}$
and $L_{2}$ are defined as
\begin{eqnarray}
&&L_{1}=\frac{1}{4}(1-s\lambda ^{2})\frac{\partial ^{2}}{\partial x^{2}}%
+s\lambda \frac{\partial }{\partial x}\frac{\partial }{\partial y}-s\frac{%
\partial ^{2}}{\partial y^{2}},  \notag \\
&&L_{2}=-\frac{1}{4}(1+s\lambda ^{2})\frac{\partial ^{2}}{\partial x^{2}}%
+s\lambda \frac{\partial }{\partial x}\frac{\partial }{\partial y}-s\frac{%
\partial ^{2}}{\partial y^{2}},  \label{maccari-1}
\end{eqnarray}%
where $\lambda \equiv K_{2}/\left( \sqrt{3}K_{1}\right) $.
This system belongs to a class of nonlinear evolution equations first studied by Shulman%
\cite{sh}, which find applications to nonlinear optics, plasma physics and
other physical fields. The integrability property and Lax pairs for system (%
\ref{maccari}) has been demonstrated by Maccari\cite{mac}. Note that, for $%
\lambda =0$, this system reduces to the Davey-Stewartson (DS) equation\cite%
{dstype}, whose integrability is a well-known fact\cite{dstype1}. Various
exact solutions have been obtained for local and nonlocal DS equations
\cite{ds1}-\cite{ds8}. Because the Maccari system (\ref{maccari}%
) is quite complex in the general form, exact solutions have been reported
for it in only in the special case of $s=-1$ \cite{lin}.

Inspired by the above considerations, we introduce an extension of system%
\eqref{maccari} satisfying the $\mathcal{PT}$ symmetry, which we refer to as
the \textit{nonlocal Maccari system}:
\begin{eqnarray}
&&iu_{t}(x,y,t)+L_{1}u(x,y,t)+u(x,y,t)v(x,y,t)=0,  \notag \\
&&L_{2}v(x,y,t)-2L_{1}[u(x,y,t)u^{\ast }(-x,-y,t)]=0,  \label{nonlocal}
\end{eqnarray}%
with the same operators $L_{1}$ and $L_{2}$ as defined by eq. %
\eqref{maccari-1}. System\eqref{nonlocal} satisfies the condition of the $2D$
$\mathcal{PT}$ symmetry. The concept of the $\mathcal{PT}$ symmetry was
first proposed by Bender and Boettcher in quantum theory\cite{bs1,bs2}. They
had shown that the spectra of non-Hermitian Hamiltonian operators are all
real as long as the $\mathcal{PT}$ symmetry is not broken. Inspired by that
pioneering work, $\mathcal{PT}$ symmetry has made a series of important
achievements in theoretical and experimental physics\cite{sy1,sy2,sy3}, and
has also driven new development in other fields, including Lie algebras \cite%
{le}, the quantum field theory\cite{lian}, complex crystals\cite{jt1,jt2},
optics and photonics. In particular, a new two-dimensional nonlinear model
was recently introduced in Ref.\cite{boris}, in which the $\mathcal{PT}$
symmetry remains unbroken for arbitrarily large values of the gain-loss
coefficient. In the same work, an
extended two-dimensional model with an imaginary part of the potential$~\sim
xy$, written in the Cartesian coordinates. The latter is not a $\mathcal{PT}$%
-symmetric model, but it also supports a continuous family of self-trapped
states, which suggests an extension of the concept of the $\mathcal{PT}$
symmetry\cite{boris}, which is another motive for addressing the nonlocal
Maccari systen\eqref{nonlocal}.

In this paper, we present various exact solutions of the nonlocal Maccari
system based on the Hirota bilinear method, long-wave limit, and the
KP-hierarchy approach. In section\ref{2}, we derive the bilinear form of the
nonlocal Maccari system, and present a new nonlocal DS-type equation. In
section\ref{3} and section\ref{4}, we obtain breather solutions on top of a
periodic background and line RWs on the periodic background, by dint of the
Hirota bilinear method and long-wave limit. In this context, semi-rational
solutions feature unique dynamics. In section\ref{5} and section\ref{6}, by
restricting different tau functions of the KP hierarchy, general line
solitons and general lump-soliton solutions are generated. In section\ref{7}
we discuss and summarize our results.


\section{Bilinear forms of nonlocal Maccari system and its reduction}\label{2}
With the help of the independent variable transformation,
\begin{equation}
u=\frac{g}{f},\quad v=\left( \frac{1}{2}(1-s\lambda ^{2})\partial
_{xx}+2s\lambda \partial _{xy}-2s\partial _{yy}\right) (\ln f),
\label{nonlocal-1}
\end{equation}%
the nonlocal Maccari system (\ref{nonlocal}) admits the following bilinear
forms:
\begin{eqnarray}
&&\left( iD_{t}+\frac{1}{4}(1-s\lambda ^{2})D_{x}^{2}+s\lambda
D_{x}D_{y}-sD_{y}^{2}\right) g(x,y,t)\cdot f(x,y,t)=0,  \notag \\
&&\left( -\frac{1}{4}(1+s\lambda ^{2})D_{x}^{2}+s\lambda
D_{x}D_{y}-sD_{y}^{2}+2\right) f\cdot f=2g(x,y,t)\cdot g^{\ast }(-x,-y,t),
\label{bilinear}
\end{eqnarray}%
where $D$ is the Hirota's bilinear differential operator\cite{hirota} defined by
\begin{equation}
\begin{aligned}
D_x^mD_t^nf(x,t)\cdot g(x,t)=(\frac{\partial}{\partial_{x}}-\frac{\partial}{\partial_{x^{'}}})^m
(\frac{\partial}{\partial_{t}}-\frac{\partial}{\partial_{t^{'}}})^nf(x,t)g(x^{'},t^{'})
\Bigg{|}_{x^{'}=x, t^{'}=t},
\end{aligned}
\end{equation}
and the following condition must hold:
\begin{equation}
f(x,y,t)=f^{\ast }(-x,-y,t).
\end{equation}

\begin{itemize}
\item When $\lambda =0$, the nonlocal Maccari system\eqref{nonlocal} reduces
to a new nonlocal DS-type system:
\begin{equation}
\left\{
\begin{array}{lr}
iu_{t}+\frac{1}{4}u_{xx}-su_{yy}+uv=0, &  \\
\frac{1}{4}v_{xx}+sv_{yy}=-\frac{1}{2}|u|_{xx}^{2}+2s|u|_{yy}^{2}=0, &  \\
|u|^{2}=u(x,y,t)u^{\ast }(-x,-y,t). &
\end{array}%
\right.  \label{ds-type}
\end{equation}

\item When $s=0$, the nonlocal Maccari system\eqref{nonlocal} reduces to the
defocusing nonlocal NLS equation:
\begin{equation}
iu_{t}(x,t)+\frac{1}{4}u_{xx}(x,t)-\frac{1}{2}u^{2}(x,t)u^{\ast }(-x,t)=0.
\label{NLS}
\end{equation}
cf. its focusing counterpart (\ref{nNLS}).
\end{itemize}

Thus, the nonlocal Maccari system\eqref{nonlocal} may be considered as a
two-dimensional generalization of the nonlocal NLS equation. Using eqs. %
\eqref{nonlocal-1} and \eqref{bilinear}, $N$-periodic wave solutions $u$ and
$v$ of the nonlocal Maccari system\eqref{nonlocal} can be obtained by means
of the Hirota method, in which $f$ and $g$ are written as follows:
\begin{equation}\label{periodic}
\begin{aligned}
&f=\sum_{\mu =0,1}\exp \left( \sum_{j<k}^{(N)}\mu _{j}\mu
_{k}A_{jk}+\sum_{j=1}^{N}\mu _{j}\eta _{j}\right) ,   \\
&g=\sum_{\mu =0,1}\exp \left( \sum_{j<k}^{(N)}\mu _{j}\mu
_{k}A_{jk}+\sum_{j=1}^{N}\mu _{j}(\eta _{j}+i\Phi _{j})\right).
\end{aligned}
\end{equation}%
Here
\begin{equation}\label{pc}
\begin{aligned}
&\eta_{j}=iP_{j}x+iq_{j}y+\Omega_{j}t+\eta^{0}_{j},
\sin(\Phi_{j})=-\frac{1}{8}\sqrt{\left[(\lambda P_j^2-2q_j)^2+P_j^2\right]^2
-2\left[(\lambda P_j^2-2q_j)^2+P_j^2\right]},\\
&\Omega_{j}=-\frac{2\sin(\Phi_{j})\left[s(2q_j-\lambda P_j)^2-P_j^2\right]}
{\left[s(2q_j-\lambda P_j)^2+P_j^2\right]},
\kappa_1=-\left[(\lambda(P_j-P_k)+2(q_k-q_j)\right]^2s+(P_j-P_k)^2,\\
&\cos(\Phi_{j})=\frac{1}{8}\left[s(2q_j-\lambda P_j)^2+P_j^2\right]+1,
\kappa_2=-\left[(\lambda(P_j+P_k)-2(q_k+q_j)\right]^2s-(P_j-P_k)^2,\\
&e^{A_{jk}}=\frac{\left[\kappa_1+4i(\Omega_{k}-\Omega_{j})\right]e^{i\Phi_{j}}
\left[\kappa_1+4i(\Omega_{j}-\Omega_{k})\right]e^{i\Phi_{k}}}
{\left[\kappa_2+4i(\Omega_{k}+\Omega_{j})\right]e^{i(\Phi_j+\Phi_k)}
\left[\kappa_2-4i(\Omega_{j}+\Omega_{k})\right]},
\end{aligned}
\end{equation}
where condition $\left[ (\lambda P_{j}^{2}-2q_{j})^{2}+P_{j}^{2}\right]
^{2}-2\left[ (\lambda P_{j}^{2}-2q_{j})^{2}+P_{j}^{2}\right] >0$ must hold.
Further, $P_{j},q_{j}$ are arbitrary real parameters, $\eta _{j}^{0}$ is a
complex constant, and $j$ is an integer subscript. The notation $\sum_{\mu
=0}$ indicates summation over all possible combinations of $\mu _{1}=0,1,\mu
_{2}=0,1,\ \cdots ,\mu _{n}=0,1$. The $\sum\limits_{j<k}^{(N)}$ summation is
running over all possible combinations of the $N$ elements.


\section{General solutions for breathers on top of the periodic line-wave
background}\label{3}
\begin{figure}[tbh]
\centering
\subfigure[t=-2]{\includegraphics[height=2.5cm,width=3.5cm]{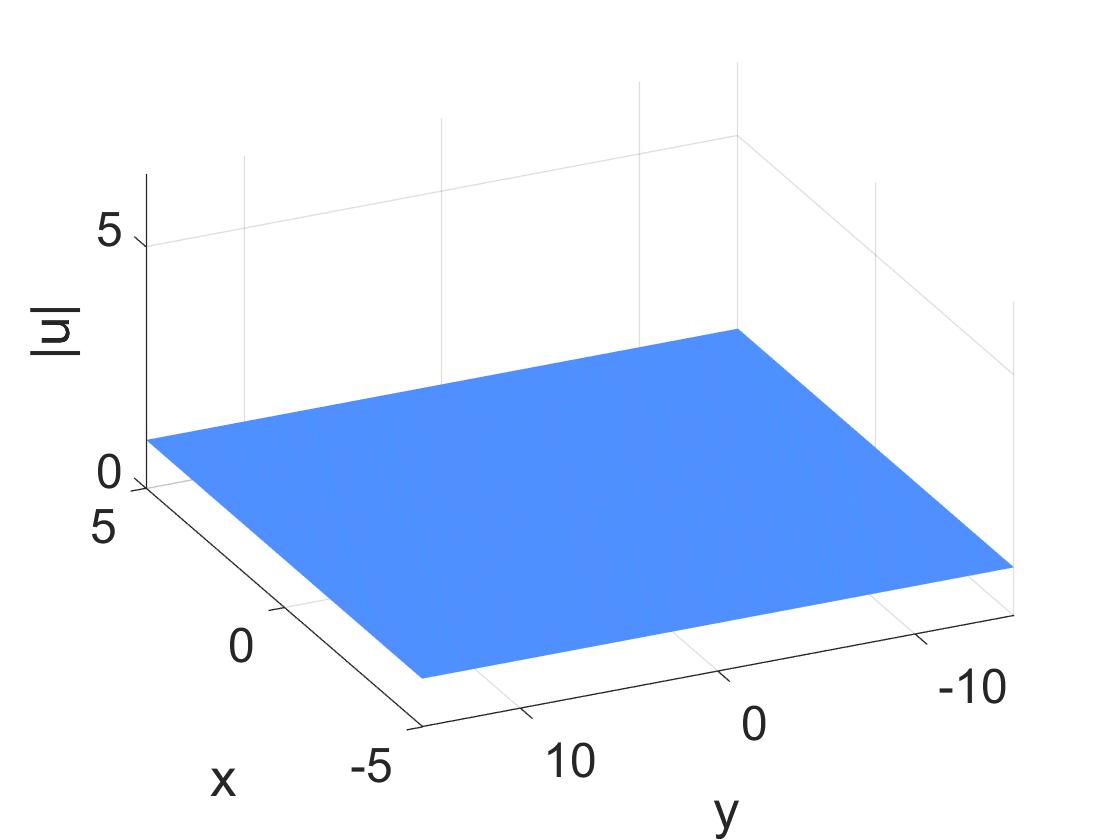}}
\subfigure[t=$\frac{1}{3}$]{\includegraphics[height=2.5cm,width=3.5cm]{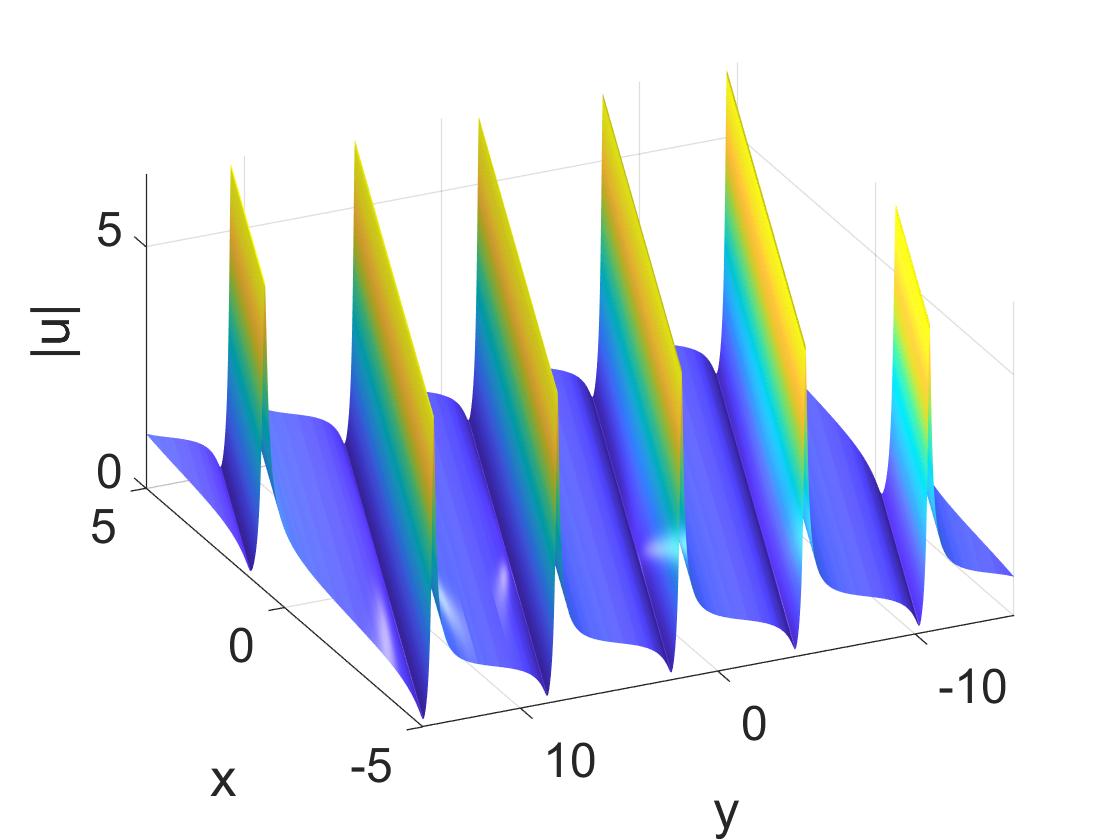}}
\subfigure[t=$\frac{1}{2}$]{\includegraphics[height=2.5cm,width=3.5cm]{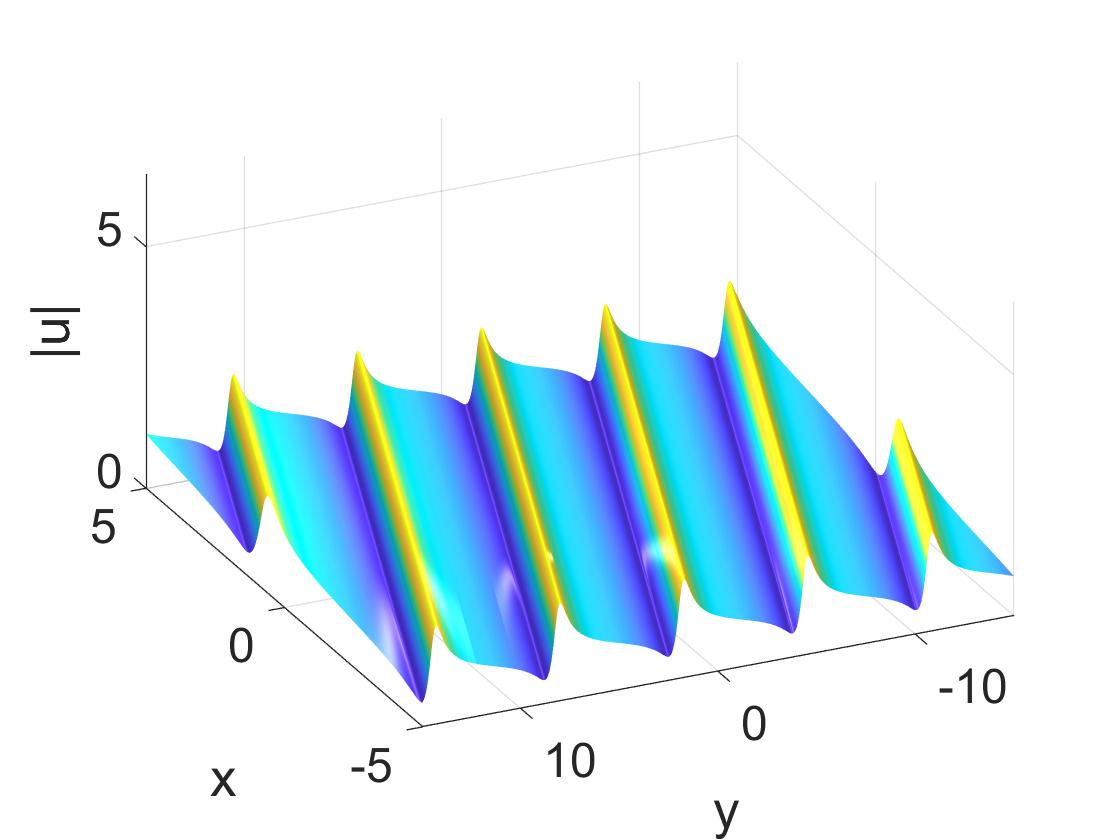}}
\subfigure[t=2]{\includegraphics[height=2.5cm,width=3.5cm]{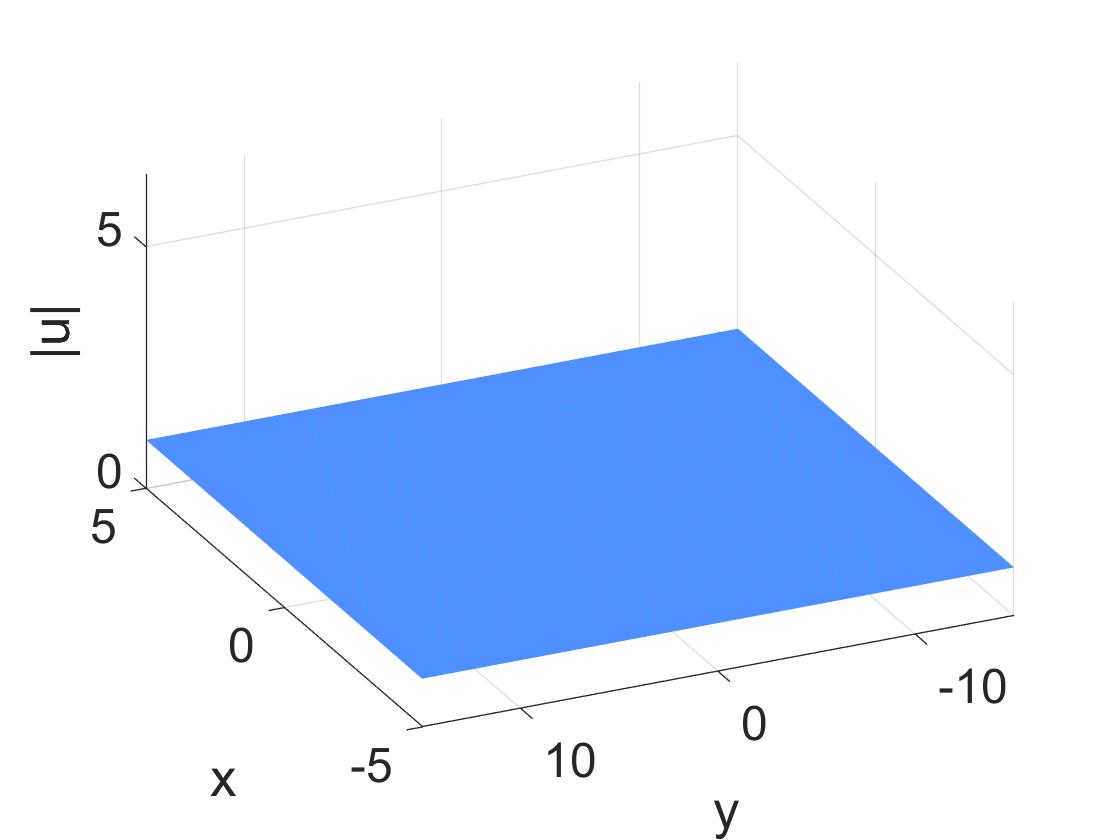}}
\caption{The evolution of the periodic line-wave solution $|u|$ of the nonlocal Maccari system.}
\label{pre}
\end{figure}

To demonstrate these periodic wave solutions, we first consider
the case of
\begin{equation*}
N=1,P_{1}=-1,q_{1}=-1,\eta _{1}^{0}=\frac{\pi }{3},\lambda =1,s=-2,
\end{equation*}%
in eq.\eqref{periodic}. The simplest periodic wave solutions $u$ and $v$
are generated as
\begin{equation}
\begin{aligned}
u=&\frac{(7+i\sqrt{15})e^{-i(x+y)-\frac{3\sqrt{15}}{4}t+\frac{\pi}{3}}+8}
{8+8e^{-i(x+y)-\frac{3\sqrt{15}}{4}t+\frac{\pi}{3}}},\\
v=&-\frac{3e^{-i(x+y)-\frac{3\sqrt{15}}{4}t+\frac{\pi}{3}}}{2(1+e^{-i(x+y)-\frac{3\sqrt{15}}{4}t+\frac{\pi}{3}})^2}.
\end{aligned}
\end{equation}
It is a series of periodic line (effectively one-dimensional) waves.
As can be seen in Fig.\ref{pre}, a series of periodic line
waves appear from the constant plane and annihilate rapidly,
and its dynamic behaviors are similar to the line breather\cite{mu3,ds3,qian,ca}.
However, the periodic line wave solution possesses one maximum amplitude
and one minimum amplitude, which is different from
the line breather with two minimum amplitudes and one
maximum amplitude.

In order to obtain breather solutions, we impose the following restrictions
in eq. \eqref{periodic}:
\begin{equation}
N=2n,P_{j}=P_{n+j},q_{j}=-q_{n+j},\eta _{j}^{0}=\eta _{n+j}^{0}.  \label{nb}
\end{equation}%
For instance, a particular case of eq. \eqref{nb}, with
\begin{equation}
N=2,P_{1}=P_{2},q_{1}=-q_{2},\eta _{j}^{0}=\eta _{n+j}^{0},  \label{1b}
\end{equation}%
is considered. Three kinds of one-breather solutions are generated by
changing the free parameter $s$, as shown seen in Fig. \ref{1b}.
\begin{figure}[tbh]
\centering
\subfigure[]{\includegraphics[height=2.5cm,width=4cm]{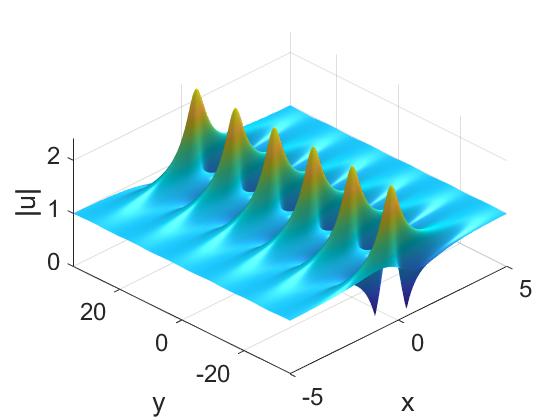}}\quad
\subfigure[]{\includegraphics[height=2.5cm,width=4cm]{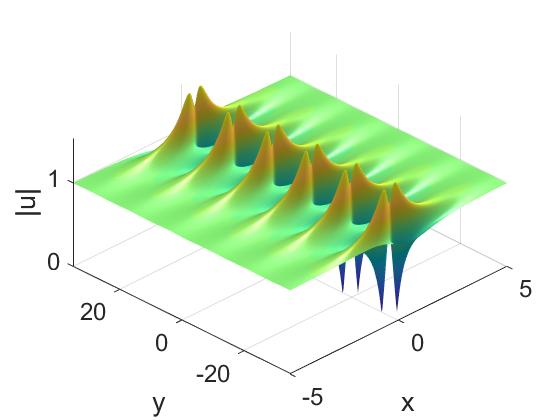}}\quad
\subfigure[]{\includegraphics[height=2.5cm,width=4cm]{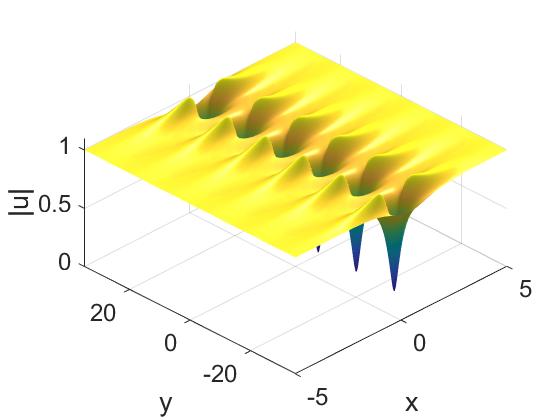}}
\caption{Three kinds of one-breather solutions of the nonlocal Maccari
system with parameters $P_{1}=i$, $P_{2}=i$, $q_{1}=\frac{1}{2}$, $q_{2}=-%
\frac{1}{2}$, $\protect\eta _{01}=\protect\eta _{02}=0$, $\protect\lambda =1$%
, displayed at $t=0$: (a): $s=-2$; (b): $s=-\frac{2}{3}$; (c): $s=-\frac{1}{5%
}$.}
\label{1b}
\end{figure}

\begin{figure}[tbh]
\centering
\subfigure[]{\includegraphics[height=2.5cm,width=4cm]{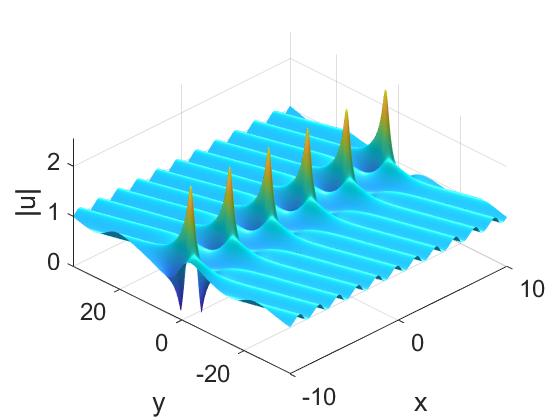}}\quad
\subfigure[]{\includegraphics[height=2.5cm,width=4cm]{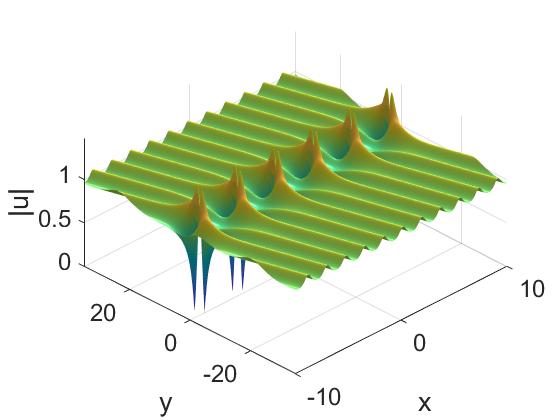}} \quad
\subfigure[]{\includegraphics[height=2.5cm,width=4cm]{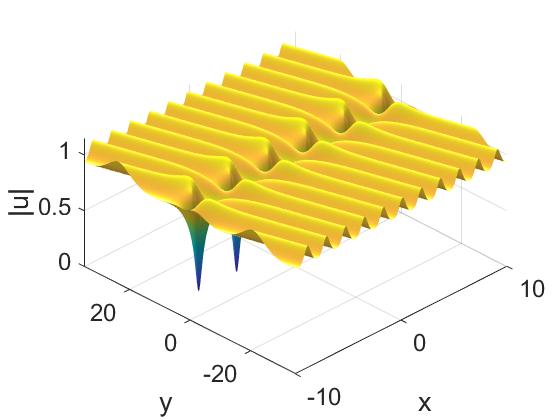}}
\caption{Three kinds of one-breather solutions on top of a periodic
line-wave background, obatined for the nonlocal Maccari system with
parameters $P_{1}=i$, $P_{2}=i$, $P_{3}=-1$, $q_{1}=\frac{1}{2}$, $q_{2}=-%
\frac{1}{2}$ $q_{3}=-1$, $\protect\eta _{01}=\protect\eta _{02}=0$, $\protect%
\eta _{03}=-\frac{\protect\pi }{2}$, $\protect\lambda =1$, displayed at $t=0$: (a): $%
s=-2$; (b): $s=-\frac{1}{2}$; (c): $s=-\frac{1}{3}$.}
\label{1bre-p}
\end{figure}

Furthermore, for the case of $N=2n+1$, a family of breathers on top of the
periodic line-wave background are presented, taking parameters
\begin{equation}
N=2n+1,P_{j}=P_{n+j},q_{j}=-q_{n+j},\eta _{j}^{0}=\eta
_{n+j}^{0},P_{2n+1}q_{2n+1}\neq 0,  \label{nbp}
\end{equation}%
in eq. \eqref{periodic}. For example, these may be
\begin{equation}
N=3,P_{1}=P_{2},q_{1}=-q_{2},P_{3}=q_{3},\eta _{1}^{0}=\eta
_{2}^{0},P_{3}q_{3}\neq 0.
\end{equation}%
The hybrid solutions, consisting of one breather and periodic line waves for
the nonlocal Maccari system, can be thus generated, see Fig. \ref{1bre-p}.

\section{General line rogue waves on top of the periodic line-wave background%
}

\label{4} In this section, we focus on RWs and semi-rational solutions of
the nonlocal Maccari system. To generate RWs, the long-wave limit may be
used. Taking parameters in eq.\eqref{periodic} as
\begin{equation}
N=2,q_{1}=\lambda _{1}P_{1},q_{2}=\lambda _{2}P_{2},\lambda _{1}=\lambda
_{2}=\kappa \neq 0,  \label{1r}
\end{equation}%
and considering the limit of $P_{j}\rightarrow 0$ $(j=1,2)$, the first-order
rational solutions $u$ and $v$ are obtained, in which $f$ and $g$ can be
expressed as
\begin{equation}\label{1r-1}
\begin{aligned}
f=&-16[(\lambda-2\kappa)^2s+1](x+\kappa y)^2+16[(\lambda-2\kappa)^2s)-1]^2t^2
+[1+(\lambda-2\kappa)^2s]^2,\\
g=&-16[(\lambda-2\kappa)^2s+1](x+\kappa y)^2+2[(\lambda-2\kappa)^2s)-1]^2
(t+\frac{-i-i(\lambda-2\kappa)^2s}{-2+2(\lambda-2\kappa)^2s}^2)^2\\
&+[1+(\lambda-2\kappa)^2s]^2.
\end{aligned}
\end{equation}
As can be seen from the above expressions, to ensure that
the rational solutions $u$ and $v$ are smooth, condition $(\lambda -2\kappa
)^{2}s+1<0$ must hold. The first-order rational solution $|u|$ is a line
rogue wave in the $(x,y)$-plane, its dynamical behavior being similar to
that of the first-order RW solution introduced in Refs.\cite{mu3,mu5,ds3,ca}%
. Additionally, higher-order RWs composed of more line RWs can also be
generated for lager $N$. Taking parameters in eq. \eqref{periodic} as
\begin{equation}
N=4,q_{j}=\lambda _{j}P_{j}(j=1,2,3,4),\lambda _{1}=\lambda _{2}\neq
0,\lambda _{3}=\lambda _{4}\neq 0,  \label{2r}
\end{equation}%
and imposing the limit of $P_{j}\rightarrow 0$ $(j=1,2)$, the second-order
rational solutions $u$ and $v$ are obtained, in which
\begin{equation}
\begin{aligned}
f=&x^4+81y^4+(\frac{321}{4}t^2-18y^2-\frac{279}{2})x^2
+(\frac{2889}{4}t^2+\frac{2673}{2})y^2+(\frac{287}{2}t^2+9)xy\\
&+\frac{105625}{72}t^2-\frac{40299}{8}t+\frac{9801}{2},\\
g=&x^4+81y^4-(18y^2+\frac{315}{2})x^2+\frac{2349}{2}y^2
-27xy+\frac{105625}{72}t^4-\frac{5525}{2}it^3+
(\frac{321}{4}x^2\\&+\frac{287}{2}xy+\frac{2889}{4}y^2
-\frac{55903}{8})t^2-(76x^2+144xy+684y^2-5355)it+\frac{12177}{2}.
\end{aligned}
\end{equation}
The corresponding solution $|u|$ is shown in Fig.\ref{2r}. It follows from
the above expressions that the solutions are subject to the boundary
condition ${\lim\limits_{x\rightarrow \pm \infty }|u|=1}$, i.e., $|u|$
approaches a constant background in the $(x,y)$-plane. A cross-shaped RW
appears at $t\approx -3/2$, describing the interaction between two line RWs.
With the increase of the interaction strength, the central region of the
cross-shaped RW creates two sharp peaks at $t\approx -3/4$. The
cross-shaped RW is separated into two hyperbolic line RWs at $t\approx 0$, whose
behavior is similar to that of the rational-soliton in the $(1+1)$-dimensional
system\cite{raso1,raso2}; then they
merge back into the constant background, the strong interaction resulting in
the change of the waveform. We stress that the dynamical behavior of the higher-order line
RWs of nonlocal Maccari system\eqref{nonlocal} is different from that of the
corresponding higher-order line RWs in other nonlocal systems, such as the
nonlocal two-dimensional NLS equation\cite{mu3,ca} and nonlocal DS equation%
\cite{s6,ds3}.
\begin{figure}[tbh]
\centering
\subfigure[t=-2]{\includegraphics[height=2.5cm,width=4cm]{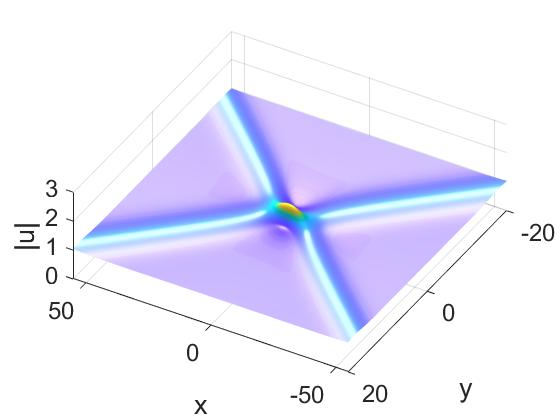}}\quad
\subfigure[t=$-\frac{3}{2}$]{\includegraphics[height=2.5cm,width=4cm]{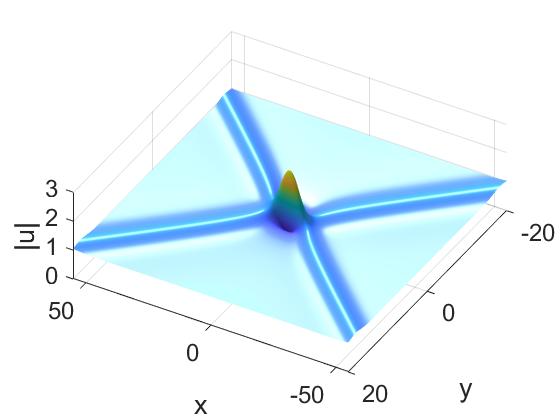}} \quad
\subfigure[t=$-\frac{3}{4}$]{\includegraphics[height=2.5cm,width=4cm]{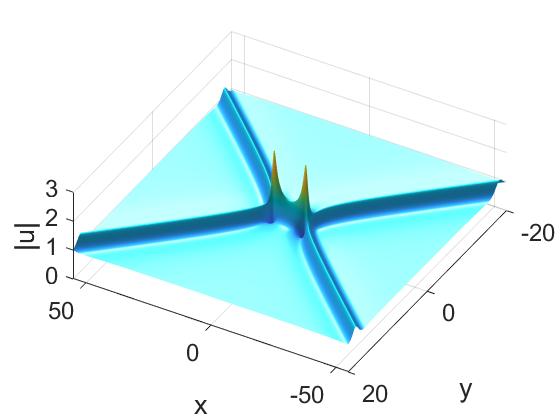}}\newline
\subfigure[t=0]{\includegraphics[height=2.5cm,width=4cm]{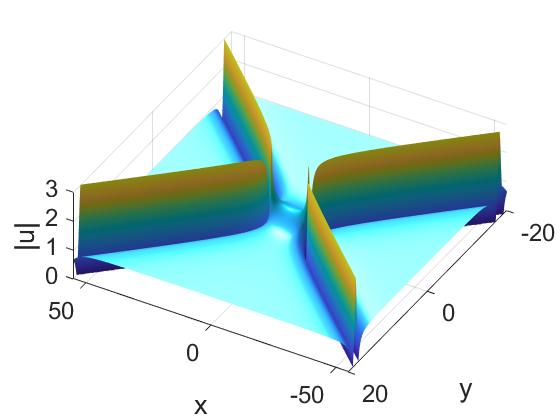}}\qquad %
\subfigure[t=2]{\includegraphics[height=2.5cm,width=4cm]{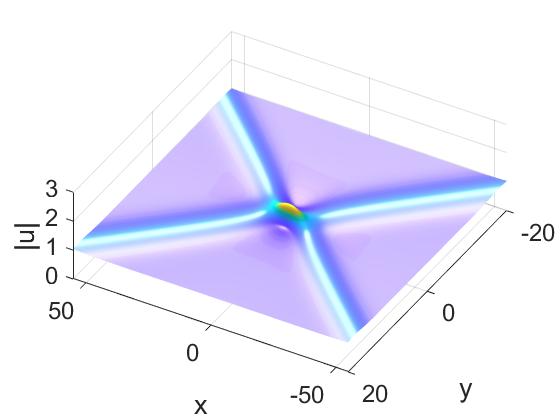}}
\caption{The temporal evolution in the $(x,y)$-plane of second-order
rogue-wave solutions of the nonlocal Maccari system with parameters $\protect%
\lambda _{1}=\protect\lambda _{2}=3$, $\protect\lambda _{3}=\protect\lambda %
_{4}=-3$, $\protect\lambda =1$ and $s=-1$.}
\label{2r}
\end{figure}

\begin{figure}[tbh]
\centering
\subfigure[t=-2]{\includegraphics[height=2.5cm,width=3.5cm]{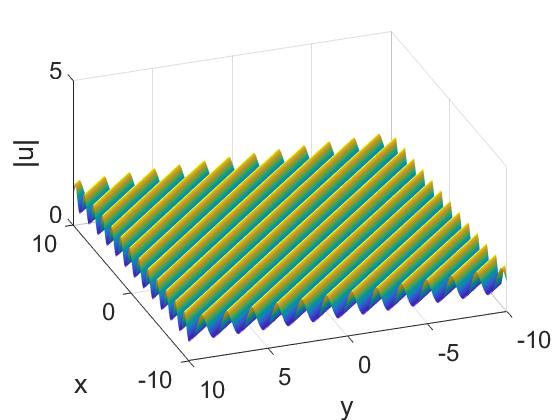}} %
\subfigure[$t=-\frac{1}{2}$]{%
\includegraphics[height=2.5cm,width=3.5cm]{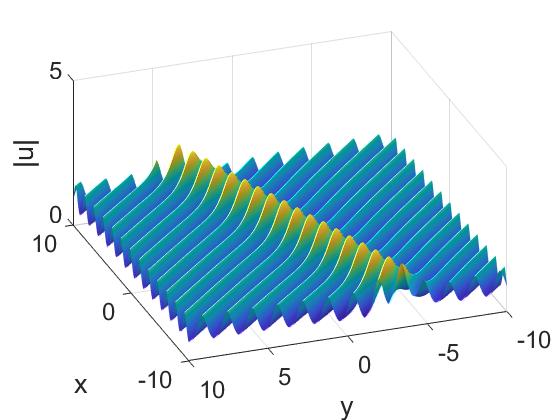}} \subfigure[t=0]{%
\includegraphics[height=2.5cm,width=3.5cm]{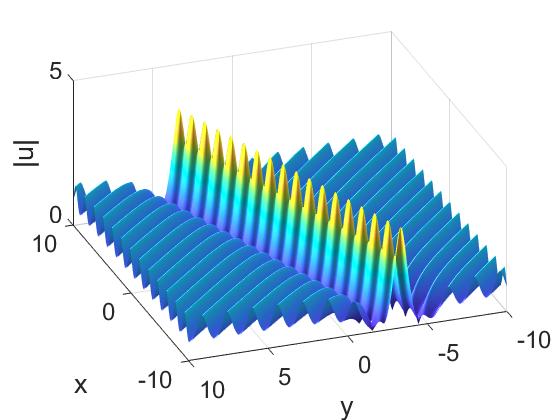}} \subfigure[t=2]{%
\includegraphics[height=2.5cm,width=3.5cm]{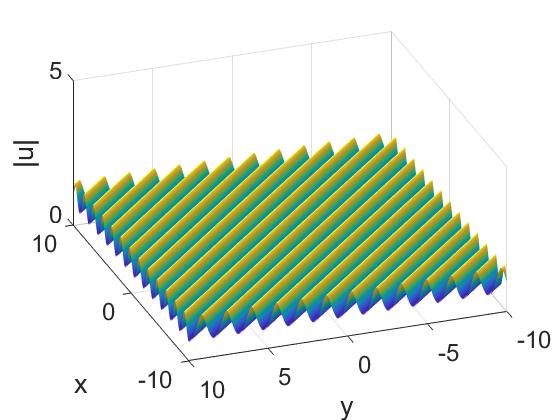}}
\caption{The evolution of the line rogue waves on top of a background of
periodic line waves, as solutions of the nonlocal Maccari system with
parameters $\protect\lambda _{1}=\protect\lambda _{2}=-3$, $P_{3}=q_{3}=4$, $%
\protect\lambda =1$ and $s=-2$.}
\label{1r1s}
\end{figure}

Semi-rational solutions of nonlocal Maccari system\eqref{nonlocal} can also
be obtained by taking the long-wave limit of a part of exponential functions
in the expressions for $f$ and $g$ given by eq. \eqref{periodic}, which
describe the interplay of RWs with the periodic waves. We first consider the
simplest semi-rational solution by setting
\begin{equation}
N=3,q_{1}=\lambda _{1}P_{1},q_{2}=\lambda _{2}P_{2},\eta _{1}^{0}=\eta
_{2}^{0}=i\pi ,  \label{N13}
\end{equation}%
and taking the limit of $P_{1},P_{2}\rightarrow 0$ in eq. \eqref{periodic},
Under parameter constraints
\begin{equation}
\lambda _{1}=\lambda _{2}=-3,P_{3}=4,q_{3}=4,\eta _{4}^{0}=-\frac{\pi }{2}%
,\lambda =1,s=-2,  \label{G3}
\end{equation}%
functions $f$ and $g$ become
\begin{equation}\label{N14}
\begin{aligned}
f=&\left(3142800(y-\frac{1}{3}x-\frac{97i}{90})^2+35283600t^2+2117025\right)e^{(-4ix-4iy-\frac{1}{2}\pi)}\\&
+3142800(y-\frac{1}{3}x)^2+35283600t^2+2117025,\\
g=&\left(-3142800(y-\frac{1}{3}x-\frac{97i}{90})^2-35283600(t-\frac{97i}{198})^2-2117025\right)e^{(-4ix-4iy-\frac{1}{2}\pi)}
\\&+349200(x-3y)^2+35283600(t-\frac{97i}{198})^2+2117025.
\end{aligned}
\end{equation}
The corresponding semi-rational solution $|u|$ is displayed in Fig.\ref{1r1s}%
. Four panels display the emergence and annihilation of a line RW on top of
the background of periodic line waves. As seen in Fig.\ref{1r1s}, a line RW
arises from the periodic line-wave background at $t\approx -1/2$, and the
interaction produces a series of sharp peaks along the line RW at $t\approx
0 $. Then, all wave peaks on the line RW ($x-3y=0$) quickly merge into the
periodic line waves. This behavior is obviously different from that featured
by the corresponding solutions of the nonlocal DS-I equation with the flat
background, cf. Ref. \cite{s6}.

Higher-order semi-rational solutions composed of two line RWs and periodic
line waves can be obtained in a similar way for $N=5$ in eq. \eqref{periodic}%
. The parameters are chosen as
\begin{equation}
N=5\,,Q_{k}=\lambda _{k}P_{k}\,,\eta _{k}^{0}=i\pi \,(\,k=1,2,3,4\,),
\label{2r1p}
\end{equation}%
with $P_{k}\rightarrow 0$ in eq. \eqref{periodic}, Under the constraints
\begin{equation}
\lambda _{1}=\lambda _{2}=-3,\lambda _{3}=\lambda _{4}=3,P_{5}=\sqrt{2}%
,q_{5}=-\sqrt{2},\eta _{5}^{0}=-\frac{\pi }{2},  \label{2r1p-1}
\end{equation}%
functions $f$ and $g$ become
\begin{equation}  \label{2r1p-2}
\begin{aligned}
f=&\{x^4+\frac{105625}{72}t^4+81y^4+(\frac{321}{4}t^2-18y^2-\frac{3168}{25})x^2
+(\frac{2889}{4}t^2+\frac{23328}{25})y^2-\frac{559623}{100}t^2\\
&+(\frac{287}{2}t^2-\frac{828}{25})xy+
i\sqrt{2}[\frac{9}{5}x^3+\frac{117}{5}x^2y-(\frac{421}{20}t^2+\frac{81}{5}y^2+\frac{2961}{25})x
-(\frac{17487}{20}t^2+\frac{1053}{5}y^2\\
&+\frac{39123}{25})y]+\frac{97119}{25}\}e^{(i\sqrt{2}x-i\sqrt{2}y-\frac{1}{2}\pi}+
x^4+81y^4+\frac{105625}{72}t^4+(\frac{321}{4}t^2-18y^2-\frac{279}{2})x^2\\
&+(\frac{2889}{4}t^2+\frac{2673}{2})y^2-\frac{40299}{8}t^2+(\frac{287}{2}t^2+9)xy+\frac{9801}{2},\\
g=\{&-x^4-81y^4-\frac{105625}{72}t^4+\frac{5525}{2}it^3-(\frac{321}{4}x^2+\frac{287}{2}xy
+\frac{2889}{4}y^2-\frac{754673}{100})t^2-\frac{19278}{25}y^2\\
&+(18y^2+\frac{3618}{25})x^2+\sqrt{2}[(\frac{421}{20}ix+\frac{17487}{20}iy)t^2
+(\frac{126}{5}x+\frac{4122}{5}y)t-\frac{9}{5}x^3-\frac{117}{5}ix^2y\\
&+(\frac{81}{5}y^2+\frac{2781}{25})ix+(\frac{1053}{5}y^2+\frac{34263}{25})iy]
+(76x^2+144xy+684y^2-\frac{146988}{25})it+\frac{1728}{25}xy\\
&-\frac{129897}{25}\}e^{(i\sqrt{2}x-i\sqrt{2}y-\frac{1}{2}\pi}
+x^4+81y^4+\frac{105625}{72}t^4-\frac{5525}{2}it^3+\frac{2349}{2}y^2
+(\frac{321}{4}x^2+\frac{287}{2}xy\\
&+\frac{2889}{4}y^2-\frac{55903}{8})t^2-(18y^2+\frac{315}{2})x^2
-(76x^2+144xy+684y^2-5355)it-27xy+\frac{12177}{2},
\end{aligned}
\end{equation}
producing semi-rational solutions $|U|$, which consist of two line RWs and
periodic line waves. As can be seen in Fig. \ref{2r1p}, two hyperbolic line
RWs appear on the periodic line-wave background at $t\approx -1/2$, and
produce a sharp peak at the top of curved RW, which seems more evident at $%
t\approx 0$. Finally, the hyperbolic RW rapidly disappears, merging into the
background of the periodic line waves. In comparison to the results reported
in Refs.\cite{s6,ds3,ca}, the corresponding solution always describes the
evolution of the cross-shaped RW on top of the background of periodic line
waves.

\begin{figure}[tbh]
\centering
\subfigure[t=-5]{\includegraphics[height=2.5cm,width=4cm]{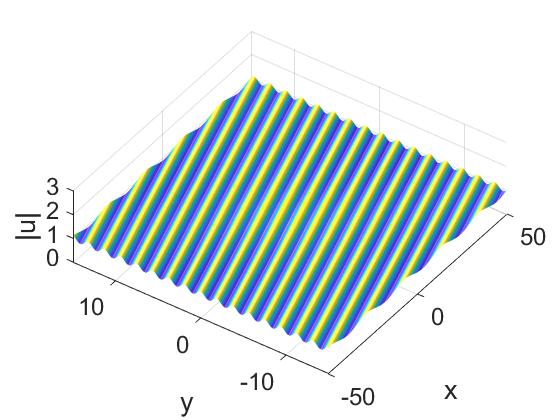}}\quad
\subfigure[t=-$\frac{1}{2}$]{\includegraphics[height=2.5cm,width=4cm]{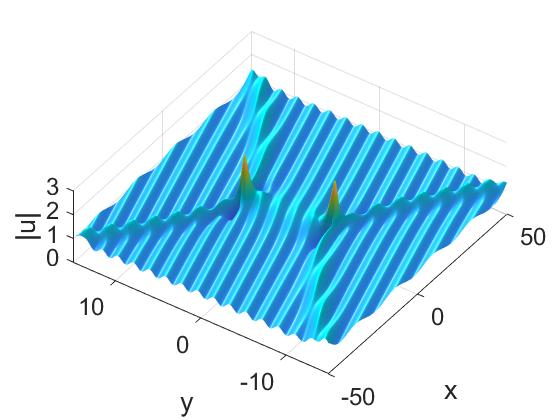}}\quad
\subfigure[t=0]{\includegraphics[height=2.5cm,width=4cm]{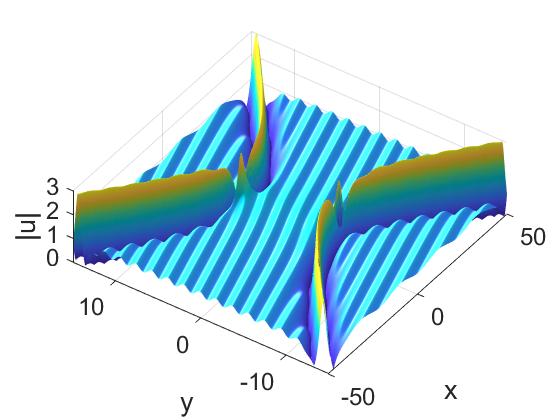}}\newline
\subfigure[t=$\frac{1}{2}$]{\includegraphics[height=2.5cm,width=4cm]{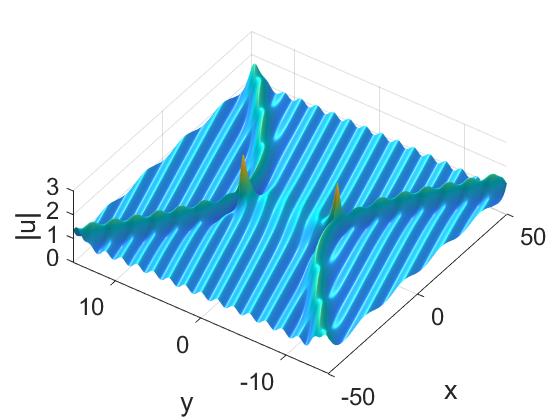}}\qquad %
\subfigure[t=5]{\includegraphics[height=2.5cm,width=4cm]{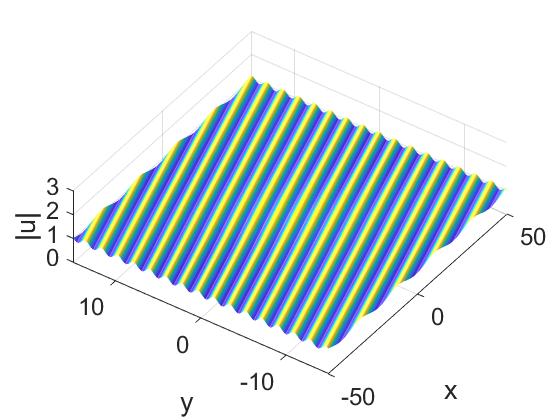}}
\caption{The evolution of the second-order line rogue waves on top of the
background of periodic line waves, obtained as a solution of the nonlocal
Maccari system with parameters $\protect\lambda _{1}=\protect\lambda _{2}=3$%
, $\protect\lambda _{3}=\protect\lambda _{4}=-3$, $P_{3}=-q_{3}=\protect%
\sqrt{2}$, $\protect\lambda =1$, and $s=-2$.}
\label{2r1p}
\end{figure}

\section{General line solitons on top of a constant background}

\label{5} In this section, we consider general soliton solutions to the
nonlocal Maccari system, built on top of a constant background. It is
difficult to get multi-soliton solutions of the nonlocal system by means of
perturbation expansion combined with the long-wave limit. Recently, general
soliton solutions for a nonlocal NLS equation were produced using a
combination of the Hirota's bilinear method and KP hierarchy reduction\cite%
{feng1}. This finding has triggered rapid progress in studies of solitons in
nonlocal systems \cite{mu4,mu5,sun1,rao1}.
Inspired by that work, we consider tau-functions of the nonlocal Maccari
system.

\noindent \textbf{Lemma 1. }Referring to the Sato theory\cite%
{ds7,sato1,sato2,sato3,sato5}, the bilinear equations in the KP hierarchy
\begin{equation}\label{kphierarchy}
\begin{aligned}
&(D_{x_{1}}^{2}-D_{x_{2}})\tau _{n+1}\cdot \tau _{n}=0,  \\
&(D_{x_{-1}}^{2}+D_{x_{-2}})\tau _{n+1}\cdot \tau _{n}=0, \\
&(D_{x_{1}}D_{x_{-1}}-2)\tau _{n}\cdot \tau _{n}=-2\tau _{n+1}\cdot \tau
_{n-1},
\end{aligned}
\end{equation}%
give rise to the following tau-functions,
\begin{equation}
\tau _{n}=\det_{1\leq j,k\leq N}(m_{j,k}^{(n)}),  \label{tau}
\end{equation}%
with matrix elements $m_{i,j}^{(n)}$ satisfying the following difference
relations,
\begin{equation}\label{tau1}
\begin{aligned}
&\partial_{x_{1}}m^{(n)}_{j,k}=\varphi^{(n)}_{j}\psi^{(n)}_{k}, \quad
m^{(n+1)}_{j,k}=m^{(n)}_{j,k}+\varphi^{(n)}_{j}\psi^{(n+1)}_{k},\\
&\partial_{x_{2}}m^{(n)}_{j,k}=\varphi^{(n+1)}_{j}\psi^{(n)}_{k}
+\varphi^{(n)}_{j}\psi^{(n-1)}_{k},\quad
\partial_{x_{-1}}m^{(n)}_{j,k}=-\varphi^{(n-1)}_{j}\psi^{(n+1)}_{k},\\
&\partial_{x_{-2}}m^{(n)}_{j,k}=-\varphi^{(n-2)}_{j}\psi^{(n+1)}_{k}
-\varphi^{(n-1)}_{j}\psi^{(n)}_{k},\\
&\partial_{x_{v}}\varphi_j^(n)=\varphi^{(n+v)}_{j}, \quad \partial_{x_{v}}\psi^(n)_{s}
=-\psi^{(n-v)}_{s} (v=-2,-1,1,2).\\
\end{aligned}
\end{equation}
Here $m_{i,j}^{(n)}$,$\varphi _{i}^{(n)}$, and $\psi _{j}^{(n)}$ are
functions of variables $x_{-1},x_{-2},x_{1}$ and $x_{2}$.

Furthermore, by changing the independent variables,
\begin{equation}
x_{-2}=-\frac{i}{2}t,x_{-1}=x+\frac{i\lambda \sqrt{s}-1}{2i\sqrt{s}}%
y,x_{1}=x+(\lambda -\frac{i\lambda \sqrt{s}-1}{2i\sqrt{s}})y,x_{2}=\frac{i}{2%
}t,  \label{bian1}
\end{equation}%
\begin{equation}
x_{-2}=-\frac{i}{2}t,x_{-1}=x+\frac{i\lambda \sqrt{s}-1}{2i\sqrt{s}}%
y,x_{1}=x+(\lambda -\frac{i\lambda \sqrt{s}+1}{2i\sqrt{s}})y,x_{2}=\frac{i}{2%
}t,
\end{equation}%
\begin{equation}
x_{-2}=-\frac{i}{2}t,x_{-1}=x+\frac{i\lambda \sqrt{s}+1}{2i\sqrt{s}}%
y,x_{1}=x+(\lambda -\frac{i\lambda \sqrt{s}-1}{2i\sqrt{s}})y,x_{2}=\frac{i}{2%
}t,
\end{equation}%
or
\begin{equation}
x_{-2}=-\frac{i}{2}t,x_{-1}=x+\frac{i\lambda \sqrt{s}+1}{2i\sqrt{s}}%
y,x_{1}=x+(\lambda -\frac{i\lambda \sqrt{s}+1}{2i\sqrt{s}})y,x_{2}=\frac{i}{2%
}t,
\end{equation}%
the bilinear equations in the KP hierarchy \eqref{kphierarchy} are reduced
to the bilinear equation \eqref{bilinear} of the nonlocal Maccari system for
\begin{equation}
f=\tau _{0},g=\tau _{1},g^{\ast }=\tau _{-1}.
\end{equation}%
Thus the nonlocal Maccari system \eqref{nonlocal} gives rise to the
following tau-functions:
\begin{equation}
u=\frac{\tau _{1}}{\tau _{0}},\quad v=\left[ \frac{1}{2}(1-s\lambda
^{2})\partial _{xx}+2s\lambda \partial _{xy}-2s\partial _{yy}\right] (\ln
\tau _{0}),
\end{equation}%
where tau-functions $\tau _{0}$ and $\tau _{1}$ satisfy relations
\begin{equation}
\tau _{0}^{\ast }(-x,-y,t)=C\tau _{0}(x,y,t),\tau _{1}^{\ast
}(-x,-y,t)=C\tau _{-1}(x,y,t).
\end{equation}%
To derive soliton solutions under nonzero boundary condition, we choose
functions $m_{j,k}^{(n)}$, $\varphi _{s}^{(n)}$ and $\psi _{s}^{(n)}$ in eq. %
\eqref{tau1} are as follows,
\begin{equation}
m_{j,k}^{(n)}=c_j\delta _{jk}+\frac{1}{p_{j}+q_{k}}\varphi _{j}^{(n)}\psi
_{k}^{(n)}, \\
\varphi _{j}^{(n)}=p_{j}^{n}e^{\xi ^{j}}, \\
\psi _{k}^{(n)}=(-q_{k})^{-n}e^{\eta _{k}},
\end{equation}%
where
\begin{equation*}
\begin{aligned}
&\xi_j=p^{-2}_jx_{-2}+p^{-1}_jx_{-1}+p_jx_1+p_j^2x_2+\xi_{j0},\\
&\eta_k=-q^{-2}_kx_{-2}+q^{-1}_kx_{-1}+q_kx_1-q_k^2x_2+\eta_{k0},
\end{aligned}
\end{equation*}
$p_{i},q_{j},\xi _{i0}$ and $\eta _{j0}$ are arbitrary complex constants,
and $\delta _{jk}$ is the Kronecker's symbol. Using the transformation of
variable given by eq. \eqref{bian1}, the above tau-function $\tau _{n}$ can
be rewritten as,
\begin{equation}
\tau _{n}=\prod_{k=1}^{N}e^{\eta _{k}+\xi _{k}}\left\vert c_{j}\delta
_{jk}e^{-(\xi _{j}+\eta _{k})}+\left( -\frac{p_{j}}{q_{k}}\right) ^{n}\frac{1%
}{p_{j}+q_{k}}\right\vert _{N\times N},  \label{lemm}
\end{equation}%
with
\begin{equation}
\begin{aligned}
&\xi_j=(p_j+\frac{1}{p_j})x+\left(\frac{\lambda(p_j^2+1)\sqrt{-s}-p_j^2+1}{2\sqrt{-s}p_j}\right)y
+\frac{i}{2}(p_j^2-\frac{1}{p_j^2})t+\xi_{j0},\\
&\eta_j=(q_j+\frac{1}{q_j})x+\left(\frac{\lambda(q_j^2+1)\sqrt{-s}-q_j^2+1}{2\sqrt{-s}q_j}\right)y
-\frac{i}{2}(q_j^2-\frac{1}{q_j^2})t+\eta_{j0}.
\end{aligned}
\end{equation}
In order to construct the line solitons on top of a constant background, the following lemma is presented.

\noindent \textbf{Lemma 2.} Setting $N=2M$, $q_{j}=p_{j}^{\ast }$, $%
c_{M+j}=-c_{j}^{\ast }$, $p_{M+j}=-p_{j}$, $\xi _{m+j0}=\xi _{j0}$, $\eta
_{m+j0}=\eta _{j0}(j=1,2,...,M)$ in eq. \eqref{lemm}, where $p_{j}$ and $%
c_{j}$ are complex parameters, $\xi _{j0}$ and $\eta _{j0}$ are real, one
has
\begin{equation*}
\tau _{n}^{\ast }(-x,-y,t)=\tau _{-n}(x,y,t),
\end{equation*}%
which means $\tau_{0}(,y,t)=f(x,y,t)$ and $\tau _{1}(,y,t)=g(x,y,t)$. A detailed proof of lemma 2 is given in Appendix A. The above results can be summarized as the following theorems.

\noindent \textbf{Theorem 1} The nonlocal Maccari system admits general
soliton solutions
\begin{equation}
u=\frac{g}{f},\quad v=\left( \frac{1}{2}(1-s\lambda ^{2})\partial
_{xx}+2s\lambda \partial _{xy}-2s\partial _{yy}\right) (\ln f),
\label{soliton}
\end{equation}%
where
\begin{equation}
\tau _{n}=\left\vert c_{j}\delta _{jk}e^{\xi _{j}+\xi _{k}^{\ast }}+\left( -%
\frac{p_{j}}{p_{k}^{\ast }}\right) ^{n}\frac{1}{p_{j}+p_{k}^{\ast }}%
\right\vert _{2M\times 2M},
\end{equation}%
\begin{equation*}
\xi _{j}=(p_{j}+\frac{1}{p_{j}})x+\left[ \frac{\lambda (p_{j}^{2}+1)\sqrt{-s}%
-p_{j}^{2}+1}{2\sqrt{-s}p_{j}}\right] y+\frac{i}{2}\left( p_{j}^{2}-\frac{1}{%
p_{j}^{2}}\right) t+\xi _{j0},
\end{equation*}%
$q_{j}=p_{j}^{\ast }$, $c_{M+j}=-c_{j}^{\ast }$, $p_{M+j}=-p_{j}$, $\xi
_{m+j0}=\xi _{j0}$, $\eta _{m+j0}=\eta _{j0}(j=1,2,...,M)$, where $p_{j}$
and $c_{j}$ are complex parameters, while $\xi _{j0}$ and $\eta _{j0}$ are
real.

\begin{figure}[tbh]
\centering
\subfigure[t=5]{\includegraphics[height=2.5cm,width=4cm]{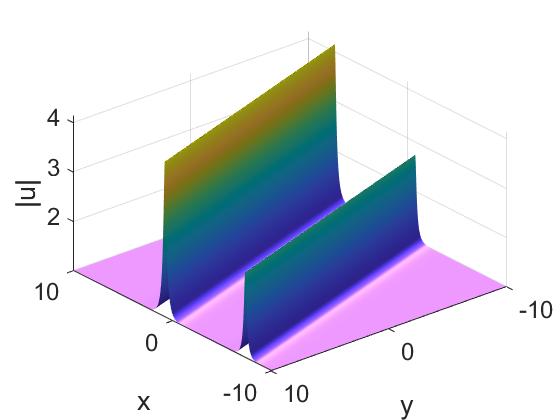}} %
\subfigure[t=5]{\includegraphics[height=2.5cm,width=4cm]{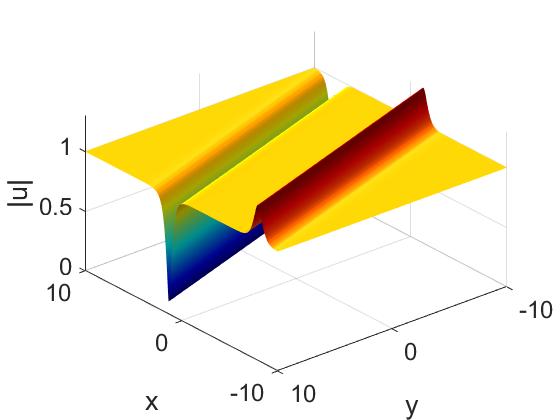}} %
\subfigure[t=5]{\includegraphics[height=2.5cm,width=4cm]{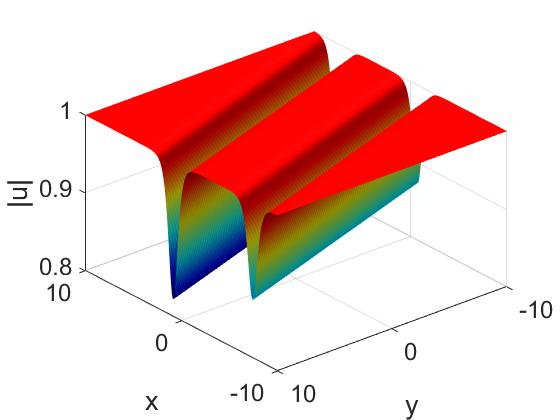}}%
\newline
\subfigure[]{\includegraphics[height=2.5cm,width=4cm]{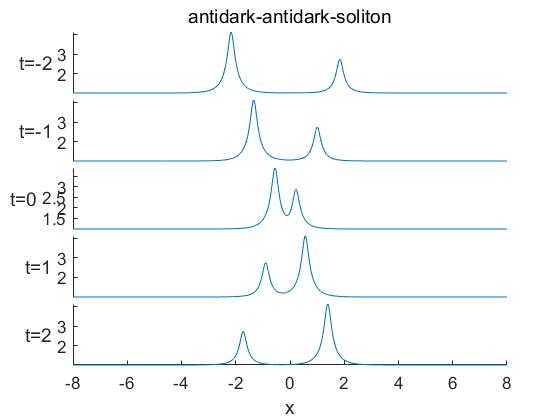}} %
\subfigure[]{\includegraphics[height=2.5cm,width=4cm]{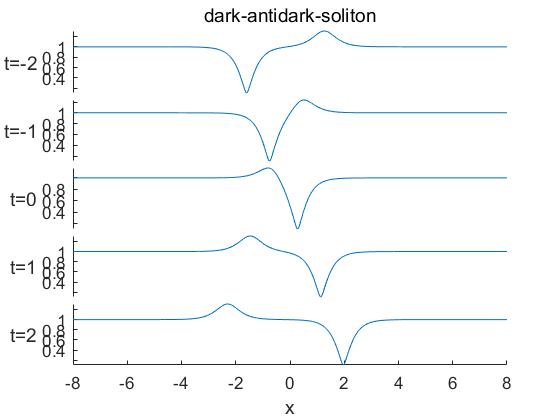}} %
\subfigure[]{\includegraphics[height=2.5cm,width=4cm]{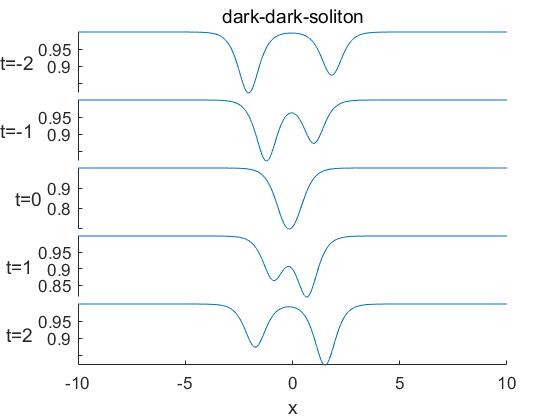}}
\caption{ Three types of two-solitons of the nonlocal Maccari system  displayed at $t=5$%
: (a) an antidark-antidark-soliton solution with parameters $p_{1}=1+i$, $%
\protect\lambda =1$, $s=-1$, and $c_{1}=-\frac{1}{3}-\frac{1}{7}i$; (b) a
dark-antidark-soliton solution with parameters $p_{1}=1+i$, $\protect\lambda %
=1$, $s=-1$, and $c_{1}=\frac{1}{3}+2i$; (c) a dark-dark-soliton solution
with parameters $p_{1}=1+i$, $\protect\lambda =1$, $s=-1$, and $c_{1}=1+%
\frac{1}{3}i$.}
\label{2soliton}
\end{figure}
In what follows, to produce soliton solutions, we first present an explicit
form of the two-soliton solutions, by setting $M=1$ in Theorem 1. The
solutions $u$ and $v$ are then expressed as
\begin{equation}
u=\frac{g}{f},\quad v=\left( \frac{1}{2}(1-s\lambda ^{2})\partial
_{xx}+2s\lambda \partial _{xy}-2s\partial _{yy}\right) (\ln f),
\label{soliton2}
\end{equation}%
where
\begin{align*}
f_{1}=&
\begin{vmatrix}
c_{1}e^{-(\xi _{1}+\xi _{1}^{\ast })}+\frac{1}{p_{1}+p_{1}^{\ast }} & \frac{1%
}{p_{1}+p_{2}^{\ast }} \\
\frac{1}{p_{2}+p_{1}^{\ast }} & -c_{1}^{\ast }e^{-(\xi _{2}+\xi _{2}^{\ast
})}+\frac{1}{p_{2}+p_{2}^{\ast }}%
\end{vmatrix}%
 \\
=& -|c_{1}|^{2}e^{i(p_{1}^{2}+p_{1}^{\ast 2}-\frac{1}{p_{1}^{2}}-\frac{1}{%
p_{1}^{\ast 2}})t}+\frac{c_{1}e^{\xi _{1}+\xi _{1}^{\ast }}}{%
p_{2}+p_{2}^{\ast }}-\frac{c_{1}^{\ast }e^{\xi _{2}+\xi _{2}^{\ast }}}{%
p_{1}+p_{1}^{\ast }} \\
& -\frac{1}{(p_{1}+p_{2}^{\ast })(p_{2}+p_{1}^{\ast })}, \\
g_{1}=&
\begin{vmatrix}
c_{1}e^{-(\xi _{1}+\xi _{1}^{\ast })}+(-\frac{p_{1}}{p_{1}^{\ast }})\frac{1}{%
p_{1}+p_{1}^{\ast }} & (-\frac{p_{1}}{p_{2}^{\ast }})\frac{1}{%
p_{1}+p_{2}^{\ast }} \\
(-\frac{p_{2}}{p_{1}^{\ast }})\frac{1}{p_{2}+p_{1}^{\ast }} & -c_{1}^{\ast
}e^{-(\xi _{2}+\xi _{2}^{\ast })}+(-\frac{p_{2}}{p_{2}^{\ast }})\frac{1}{%
p_{2}+p_{2}^{\ast }}%
\end{vmatrix}%
 \\
=& -|c_{1}|^{2}e^{i(p_{1}^{2}+p_{1}^{\ast 2}-\frac{1}{p_{1}^{2}}-\frac{1}{%
p_{1}^{\ast 2}})t}-\frac{c_{1}e^{\xi _{1}+\xi _{1}^{\ast }}}{%
p_{2}+p_{2}^{\ast }}(-\frac{p_{2}}{p_{2}^{\ast }})+\frac{c_{1}^{\ast }e^{\xi
_{2}+\xi _{2}^{\ast }}}{p_{1}+p_{1}^{\ast }}(-\frac{p_{1}}{p_{1}^{\ast }}) \\
& -\frac{1}{(p_{1}+p_{2}^{\ast })(p_{2}+p_{1}^{\ast })}(-\frac{p_{1}p_{2}}{%
p_{1}^{\ast }p_{2}^{\ast }}),
\end{align*}%
and
\begin{equation}
\begin{aligned}
\xi_1+\eta^*_1=&(p_1+p_1^*+\frac{1}{p_1}+\frac{1}{p^*_1})x+
\left(\frac{\lambda(p_1^2+1)\sqrt{-s}-p_1^2+1}{2\sqrt{-s}p_1}
+\frac{\lambda(p_1^{*2}+1)\sqrt{-s}-p_1^{*2}+1}{2\sqrt{-s}p_1^*}\right)y\\
&+\frac{i}{2}(p_1^2-\frac{1}{p_1^2}+p_1^{*2}-\frac{1}{p_1^{*2}})t+\xi_{10}+\eta_{10},\\
\xi_2+\eta^*_2=&-(\xi_1+\eta^*_1)+i(p_1^2-\frac{1}{p_1^2}+p_1^{*2}-\frac{1}{p_1^{*2}})t.
\end{aligned}
\end{equation}
It can be seen from the above expressions that the two-soliton solution is
controlled by two free parameters, $p_{1}$ and $c_{1}$. There are three
patterns of two-soliton solutions for a given parameter $p_{1}$. When $%
c_{1}=-\frac{1}{3}-\frac{1}{7}i$, the corresponding solutions are two
anti-dark solitons; when $c_{1}=\frac{1}{3}+2i$, the solutions are
dark-anti-dark solitons; when $c_{1}=1+\frac{1}{3}i$, the solutions are two
dark solitons, see Fig. \ref{2soliton}(a,b,c). We stress that the two
solitons form a parallel pair, and their dynamical behavior is different
from that of the cross solitons in nonlocal systems considered in Refs. \cite%
{mu4,mu5,sun1}. As can be seen in Fig. \ref{2soliton}(d,e,f), the two
solitons pass through each other without changes in their velocity and
waveforms, which suggests that there the interaction between the two
solitons is strictly elastic, as in other integrable systems.

\begin{figure}[tbh]
\centering
\subfigure[]{\includegraphics[height=2.5cm,width=5cm]{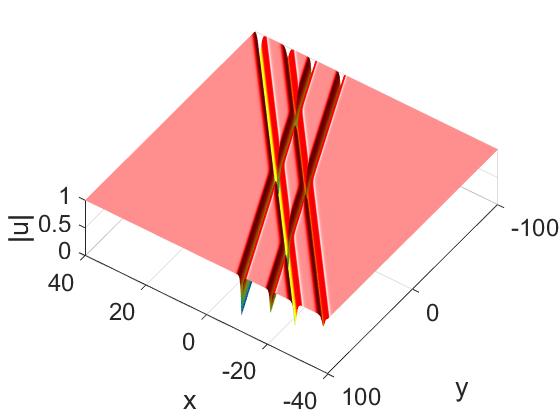}} %
\subfigure[]{\includegraphics[height=2.5cm,width=5cm]{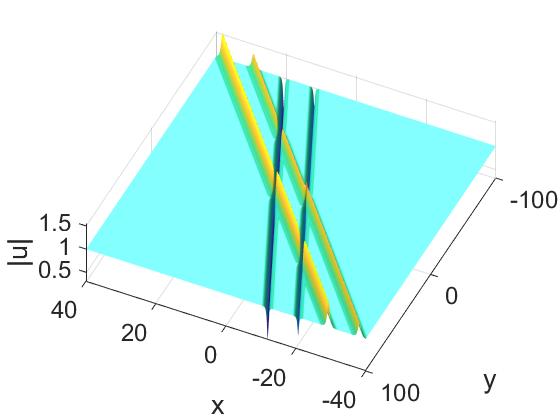}} %
\subfigure[]{\includegraphics[height=2.5cm,width=5cm]{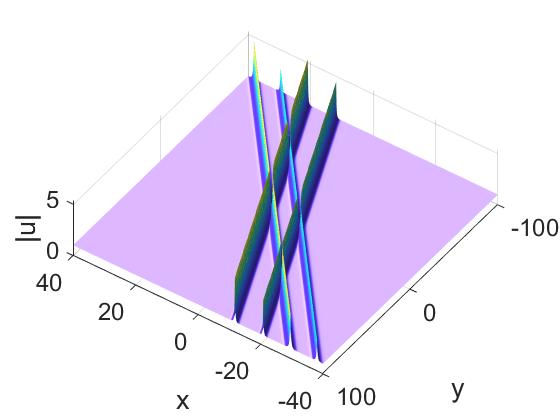}}\newline
\subfigure[]{\includegraphics[height=2.5cm,width=4cm]{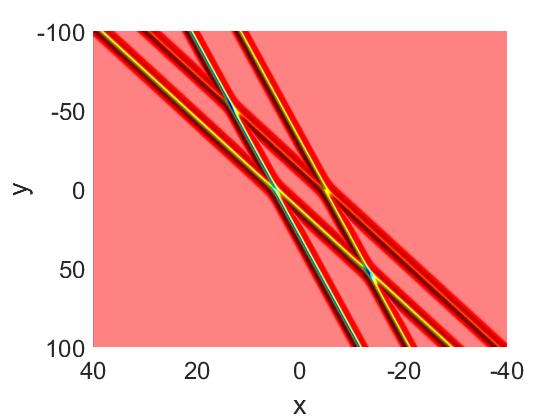}} %
\subfigure[]{\includegraphics[height=2.5cm,width=4cm]{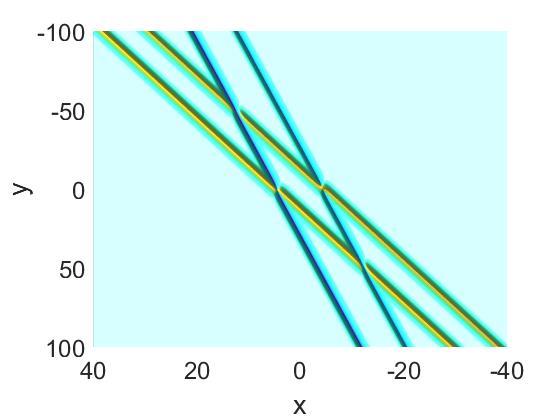}} %
\subfigure[]{\includegraphics[height=2.5cm,width=4cm]{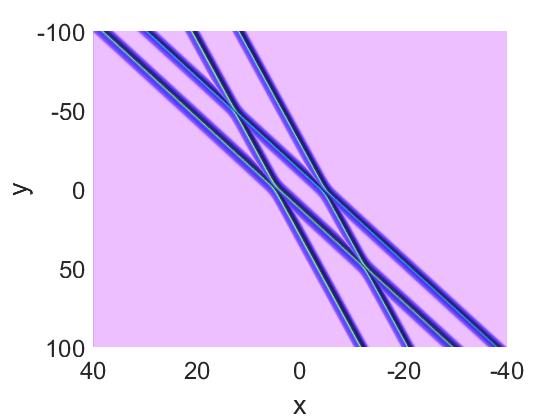}}
\caption{Three types of four-soliton solutions of the nonlocal Maccari
system displayed at $t=5$: (a) four antidark solitons with parameters $%
p_{1}=1+i$, $p_{2}=2-i$, $\protect\lambda =1$, $s=-2$, $c_{1}=-\frac{1}{3}-%
\frac{1}{8}i$, and $c_{2}=-\frac{1}{3}-\frac{1}{8}i$; (b) a
two-dark-two-antidark-solitons complex with parameters $p_{1}=1+i$, $%
p_{2}=2-i$, $\protect\lambda =1$, $s=-2$, $c_{1}=-\frac{1}{3}-2i$, and $%
c_{2}=-\frac{1}{3}-2i$; (c) four dark solitons with parameters $p_{1}=1+i$, $%
p_{2}=2-i$, $\protect\lambda =1$, $s=-2$, $c_{1}=1+\frac{1}{3}i$, and $%
c_{2}=1-\frac{1}{3}i$.}
\label{4soliton}
\end{figure}

Furthermore, solutions for collisions of $2M$ line solitons can be found for
larger $M$. The patterns of $2M$ line solitons are controlled by parameters $%
p_{k}$ and $c_{k}$ ($k=1,2,\cdot \cdot \cdot ,M$). For example, taking $M=2$%
, three patterns of four-soliton interactions are presented for given
parameters $p_{1}$ and $p_{2}$ in Fig. \ref{4soliton}.


\section{General lumps on solitons background}\label{6}
In this section, we focus on semi-rational solutions produced by
dint of the KP-hierarchy method. To obtain semi-rational solutions of the
nonlocal Maccari system \eqref{nonlocal}, we introduce the following
differential operators\cite{ds7,sato5}:
\begin{equation}
\Xi _{s}=\sum_{k=0}^{n_{0}}c_{sk}(p_{s}\partial _{p_{s}})^{n_{0}-k},\quad
\mho _{j}=\sum_{l=0}^{n_{0}}d_{jl}(q_{j}\partial _{q_{j}})^{n_{0}-l},
\end{equation}%
and choose the following functions,
\begin{equation}
\begin{aligned}
&\varphi^{(n)}_{s}=\Xi_sp^{n}_{s}e^{\xi^{s}},\\
&\psi^{(n)}_{j}=\mho_j(-q_{j})^{-n}e^{\eta_{j}},\\
&M^{(n)}_{s,j}=\Xi_s\mho_j\frac{1}{p_{s}+q_{j}}[\delta_{sj}
+(-\frac{p_s}{q_j})^ne^{\xi_s+\eta_j}].\\
\end{aligned}
\end{equation}
Functions $\varphi _{s}^{(n)}$ and $\psi _{j}^{(n)}$ also satisfy eq. %
\eqref{tau1}. For simplicity, we rewrite matrix element $M_{s,j}^{(n)}$ as \cite{ds7,sato5}
\begin{equation}\label{semi}
\begin{aligned}
M_{s,j}^{(n)}&=(-\frac{p_s}{q_j})^ne^{\xi_s+\eta_j}\sum_{k=0}^{n_0}a_{sk}
(p_s\partial_{p_s}+\xi^{'}_s+n)^{n_0-k}\\
& \times \sum_{l=0}^{n_0}d_{jl}(q_j
\partial_{q_j}+\eta^{'}_j-n)^{n_0-l}\frac{1}{p_s+q_j}+\widetilde{c}_{sj},
\end{aligned}
\end{equation}
where
\begin{equation}
\begin{aligned}
&\xi_j=(p_j+\frac{1}{p_j})x+\left(\frac{\lambda(p_j^2+1)\sqrt{-s}-p_j^2+1}{2\sqrt{-s}p_j}\right)y
+\frac{i}{2}(p_j^2-\frac{1}{p_j^2})t+p_j\widetilde{\xi}_{j},\\
&\eta_j=(q_j+\frac{1}{q_j})x+\left(\frac{\lambda(q_j^2+1)\sqrt{-s}-q_j^2+1}{2\sqrt{-s}q_j}\right)y
-\frac{i}{2}(q_j^2-\frac{1}{q_j^2})t+q_j\widetilde{\eta}_{j},\\
&\xi^{'}_j=(p_j-\frac{1}{p_j})x+\left(\frac{\lambda(p_j^2-1)\sqrt{-s}-p_j^2-1}{2\sqrt{-s}p_j}\right)y
+i(p_j^2+\frac{1}{p_j^2})t+p_j\widetilde{\xi}_{j},\\
&\eta^{'}_j=(q_j-\frac{1}{q_j})x+\left(\frac{\lambda(q_j^2-1)\sqrt{-s}-q_j^2-1}{2\sqrt{-s}q_j}\right)y
-i(q_j^2+\frac{1}{q_j^2})t+q_j\widetilde{\eta}_{j},
\end{aligned}
\end{equation}
Here $p_{i}$ and $a_{ik}$ are arbitrary complex constants, $\delta _{ij}=0,1$%
, $n_{i}$ are arbitrary positive integers, and $\widetilde{c}_{sj}=\delta
_{sj}c_{s,n_{s}}c_{j,n_{j}}^{\ast }$.

\noindent\textbf{Lemma 3} Consider the $2N\times2N$ matrix for the tau function
$M^{(n)}_{s,j}$ defined in \eqref{semi}. Taking
\begin{equation}
\begin{aligned}
&p_{N+j}=-p_j, \widetilde{c}_{N+s,N+j}=-\widetilde{c}^*_{j,s},
\widetilde{c}_{j,s}=\widetilde{c}_{s,j}, \widetilde{c}_{N+s,j}=ib_{N+s,j},
\widetilde{c}_{s,N+j}=ib_{s,N+j},\\
&b_{s,N+j}=b_{j,N+s}, c_{N+s,j}=c_{s,j}, c_{s,N+j}=c_{s,j},
\widetilde{\eta}_{j}=-\widetilde{\xi}^*_{j},\widetilde{\xi}_{N+s}=\widetilde{\xi}_{s},
q_j=p^*_j,c^*_{sk}=d_{sk},
\end{aligned}
\end{equation}
for $s,j=1,2,\cdot\cdot\cdot,N$, then we have
$$\tau^*_n(-x,-y,t)=\tau_{-n}(x,y,t).$$
The proof of lemma 3 is given in Appendix B. According to Lemma 3,
the semi-rational solutions of the nonlocal Maccari system \eqref{nonlocal}
can be summarized in the following Theorem.

\noindent \textbf{Theorem 2.} The nonlocal Maccari system \eqref{nonlocal}
has semi-rational solutions
\begin{equation}
u=\frac{\tau _{1}}{\tau _{0}},\quad v=\left( \frac{1}{2}(1-s\lambda
^{2})\partial _{xx}+2s\lambda \partial _{xy}-2s\partial _{yy}\right) (\ln
\tau _{0}),  \label{theor}
\end{equation}%
where
\begin{equation}
\tau _{n}(x,y,t)=%
\begin{vmatrix}
M_{j,s}^{(n)} & M_{N+j,s}^{(n)} \\
M_{j,N+s}^{(n)} & M_{N+j,N+s}^{(n)}%
\end{vmatrix}%
_{(0<s,j \le N)},
\end{equation}%
and the matrix elements are given by
\begin{equation}\label{xuyao}
\begin{aligned}
M_{j,s}^{(n)}&=(-\frac{p_j}{p^*_s})^ne^{\xi_j+\xi^*_s}\sum_{k=0}^{n_0}c_{jk}
(p_j\partial_{p_j}+\xi^{'}_j+n)^{n_0-k}\\
& \times \sum_{l=0}^{n_0}c^*_{sl}(p^*_s
\partial_{p^*_s}+\xi^{'*}_j-n)^{n_0-l}\frac{1}{p_j+p^*_s}+\widetilde{c}_{sj},
\end{aligned}
\end{equation}
with
\begin{equation}\label{para}
\begin{aligned}
&\xi_j=(p_j+\frac{1}{p_j})x+\left(\frac{\lambda(p_j^2+1)\sqrt{-s}-p_j^2+1}{2\sqrt{-s}p_j}\right)y
+\frac{i}{2}(p_j^2-\frac{1}{p_j^2})t+p_j\widetilde{\xi}_{j},\\
&\xi^{'}_j=(p_j-\frac{1}{p_j})x+\left(\frac{\lambda(p_j^2-1)\sqrt{-s}-p_j^2-1}{2\sqrt{-s}p_j}\right)y
+i(p_j^2+\frac{1}{p_j^2})t+p_j\widetilde{\xi}_{j},
\end{aligned}
\end{equation}
and parameters satisfying the following equations:
\begin{equation}
\begin{aligned}
&p_{N+j}=-p_j, \widetilde{c}_{N+s,N+j}=-\widetilde{c}^*_{j,s},
\widetilde{c}_{j,s}=\widetilde{c}_{s,j}, \widetilde{c}_{N+s,j}=ib_{N+s,j},
\widetilde{c}_{s,N+j}=ib_{s,N+j},\\
&b_{s,N+j}=b_{j,N+s}, c_{N+s,j}=c_{s,j}, c_{s,N+j}=c_{s,j},
\widetilde{\eta}_{j}=-\widetilde{\xi}^*_{j},\widetilde{\xi}_{N+s}=\widetilde{\xi}_{s}.
q_j=p^*_j,c^*_{sk}=d_{sk}.
\end{aligned}
\end{equation}

\noindent \textbf{Remark 1:} Setting $\widetilde{c}_{sj}=0$ and $%
b_{N+s,j}=b_{s,N+j}=0(1\leq s,j\leq N)$ reduces the semi-rational solutions %
\eqref{theor} to rational solutions of the nonlocal Maccari system %
\eqref{nonlocal}, which are $2N$-lump-type solutions.

\noindent \textbf{Remark 2:} With $\widetilde{c}_{sj}\neq 0$, these
semi-rational solutions \eqref{theor} describe the interaction between $2N$%
-lumps and $2N$-solitons of the nonlocal Maccari system \eqref{nonlocal}.

\subsection{General lumps solutions}
\begin{figure}[tbh]
\centering
\subfigure[]{\includegraphics[height=2.5cm,width=5cm]{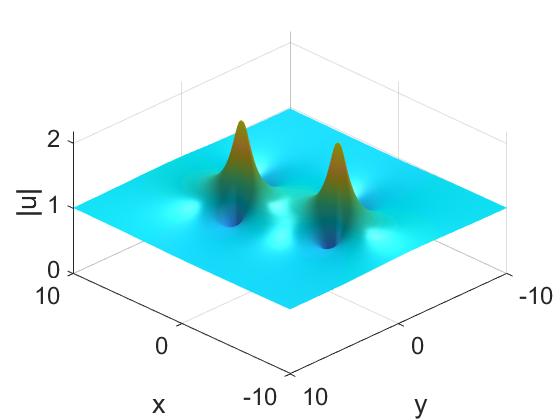}} %
\subfigure[]{\includegraphics[height=2.5cm,width=5cm]{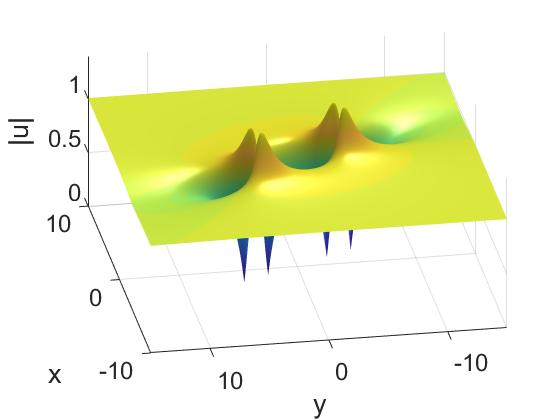}} %
\subfigure[]{\includegraphics[height=2.5cm,width=5cm]{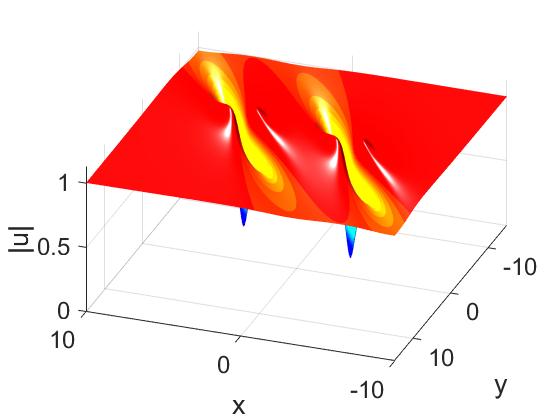}}\newline
\subfigure[]{\includegraphics[height=2.5cm,width=4cm]{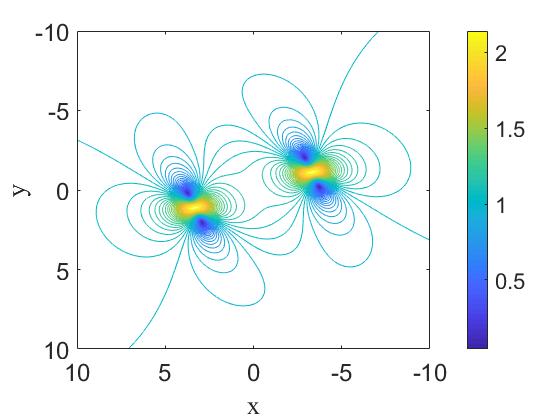}} %
\subfigure[]{\includegraphics[height=2.5cm,width=4cm]{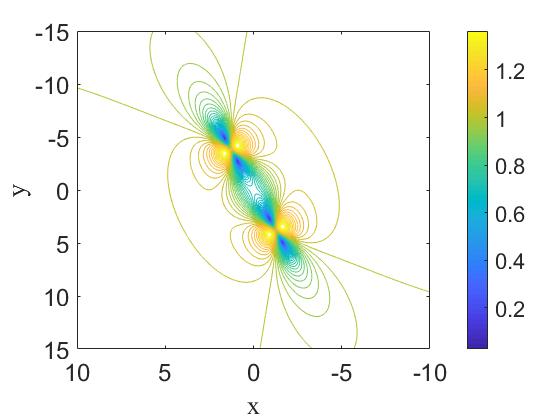}} %
\subfigure[]{\includegraphics[height=2.5cm,width=4cm]{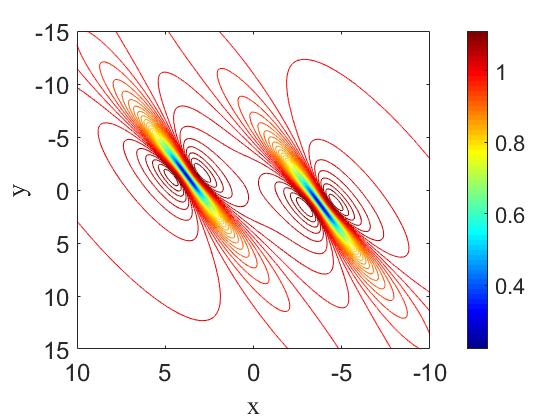}}
\caption{Three species of two-lump solutions of the nonlocal Maccari
equation are displayed at $t=2$: (a) a bright-bright-lump solution for
parameters $p_{1}=1+\frac{1}{2}i$, $\protect\lambda =1$, and $s=-1$; (b) a
four-petal-four-petal-lump solution for parameters $p_{1}=1+i$, $\protect%
\lambda =1$, and $s=-1$; (c) a dark-dark-lump solution for parameters $%
p_{1}=1+2i$, $\protect\lambda =1$, and $s=-1$.}
\label{2lump}
\end{figure}
To illustrate the $2N$-lump solutons, we first take $N=1,n_{0}=1,\widetilde{c%
}_{11}=0$ and $b_{12}=0$ in eq. \eqref{theor}, we obtain
\begin{equation}
u=\frac{g}{f},\quad v=\left[ \frac{1}{2}(1-s\lambda ^{2})\partial
_{xx}+2s\lambda \partial _{xy}-2s\partial _{yy}\right] (\ln f),
\end{equation}%
where
\begin{equation}
\begin{aligned}
f&=\begin{vmatrix}
\zeta_1\zeta_1^{'}+\frac{p_1p_1^{*}}{(p_1+p_1^{*})^2} & \zeta_1\zeta_2^{'}+\frac{p_1p_2^{*}}{(p_1+p_2^{*})^2}
\\ \zeta_2\zeta_1^{'}+\frac{p_2p_1^{*}}{(p_2+p_1^{*})^2}& \zeta_2\zeta_2^{'}+\frac{p_2p_2^{*}}{(p_2+p_2^{*})^2} \end{vmatrix},\\
g&=\begin{vmatrix}
(\zeta_1+1)(\zeta_1^{'}-1)+\frac{p_1p_1^{*}}{(p_1+p_1^{*})^2}& (\zeta_1+1)(\zeta_2^{'}-1)+\frac{p_1p_2^{*}}{(p_1+p_2^{*})^2}
 \\(\zeta_2+1)(\zeta_1^{'}-1)+\frac{p_2p_1^{*}}{(p_2+p_1^{*})^2} & (\zeta_2+1)(\zeta_2^{'}-1)+\frac{p_2p_2^{*}}{(p_2+p_2^{*})^2}
 \end{vmatrix},\\
 \end{aligned}
\end{equation}
with
\begin{equation}\label{zeta}
\begin{aligned}
\zeta_s=\xi^{'}_s-\frac{p_s}{p_s+p^{*}_j}+c_{11},\zeta^{'}_s=\eta^{'}_s-\frac{p^{*}_j}{p^{*}_j+p_s}+c^{*}_{11},
\end{aligned}
\end{equation}
where $\xi ^{_{s}^{{\prime }}}$ and $\eta ^{_{j}^{{\prime }}}$ $(s,j=1,2)$
are defined in eq. \eqref{para}, $p_{2}=-p_{1}$, $\widetilde{\xi }_{2}=-%
\widetilde{\xi }_{1}$, $p_{1}$, $c_{11}$, $\widetilde{\xi }_{1}$ being
complex parameters. The corresponding rational solutions are two lumps in
the $(x,y)$-plane, whose shape is controlled by parameters $p_{1}$ and $%
\widetilde{\xi }_{1}$. When $\widetilde{\xi }_{1}=0$, as discussed in Ref.
\cite{ctp}, the shape of each lump is controlled by parameters $\frac{p_{1Re}%
}{p_{1Im}}$: a dark lump for $0<\frac{p_{1Re}^{2}}{p_{1Im}^{2}}\leq \frac{1%
}{3}$; a four-petal lump for $\frac{1}{3}<\frac{p_{1Re}^{2}}{p_{1Im}^{2}}%
\leq 3$; and a bright lump for $\frac{p_{1Re}^{2}}{p_{1Im}^{2}}>3$. Here $%
p_{1Re}$ and $p_{1Im}$ are the real part and the imaginary part of $p_{1}$.
As $\frac{p_{1Re}^{2}}{p_{1Im}^{2}}=\frac{p_{2Re}^{2}}{p_{2Im}^{2}}$, two
lumps have the same structure. Three species of two-lumps patterns, namely,
two bright lumps, two four-petal lumps, and two dark lumps, are shown in
Fig. \ref{2lump}.

Similarly, four-lump solutions of the nonlocal Maccari system %
\eqref{nonlocal} are generated by setting
\begin{equation*}
N=2,n_{0}=1,\widetilde{c}_{s,j}=0,b_{s,j}=0,c_{s,0}=1,(s\neq j,s,j=1,2,3,4),
\end{equation*}%
in eq. \eqref{theor}, the corresponding solutions $u$ and $v$ being
\begin{equation}
u=\frac{g}{f},\quad v=\left[ \frac{1}{2}(1-s\lambda ^{2})\partial
_{xx}+2s\lambda \partial _{xy}-2s\partial _{yy}\right] (\ln f),
\end{equation}%
with
\begin{equation}\label{soliton}
\begin{aligned}
f(x,y,t)&=\begin{vmatrix} M_{11}^{(0)} & M_{12}^{(0)} & M_{13}^{(0)} &M_{14}^{(0)}
 \\ M_{21}^{(0)} & M_{22}^{(0)} & M_{23}^{(0)} & M_{24}^{(0)} \\
M_{31}^{(0)} & M_{32}^{(0)} & M_{33}^{(0)} & M_{34}^{(0)} \\
M_{41}^{(0)} & M_{42}^{(0)} & M_{43}^{(0)} & M_{44}^{(0)} \end{vmatrix},\\
g(x,y,t)&=\begin{vmatrix} M_{11}^{(1)} & M_{12}^{(1)} & M_{13}^{(1)} &M_{14}^{(1)}
 \\ M_{21}^{(1)} & M_{22}^{(1)} & M_{23}^{(1)} & M_{24}^{(1)} \\
M_{31}^{(1)} & M_{32}^{(1)} & M_{33}^{(1)} & M_{34}^{(1)} \\
M_{41}^{(1)} & M_{42}^{(1)} & M_{43}^{(1)} & M_{44}^{(1)} \end{vmatrix},
\end{aligned}
\end{equation}
where $M_{s,j}^{(n)}(s,j=1,2,3,4,n=0,1)$ are defined in eq. \eqref{xuyao}.
Three species of four-lump solutions, namely, four bright lumps,
four-lump-four-petal lumps, and four dark lumps, are displayed in Fig. \ref%
{4lump}.
\begin{figure}[tbh]
\centering
\subfigure[]{\includegraphics[height=2.5cm,width=5cm]{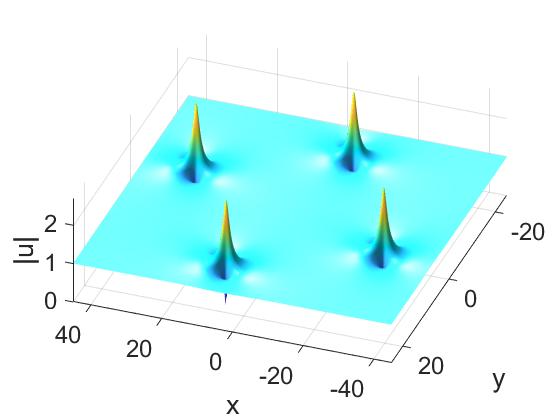}} %
\subfigure[]{\includegraphics[height=2.5cm,width=5cm]{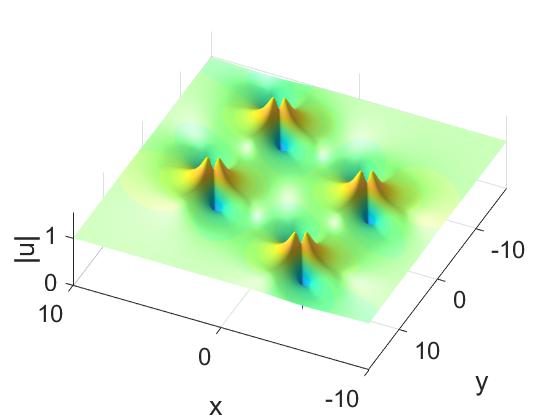}} %
\subfigure[]{\includegraphics[height=2.5cm,width=5cm]{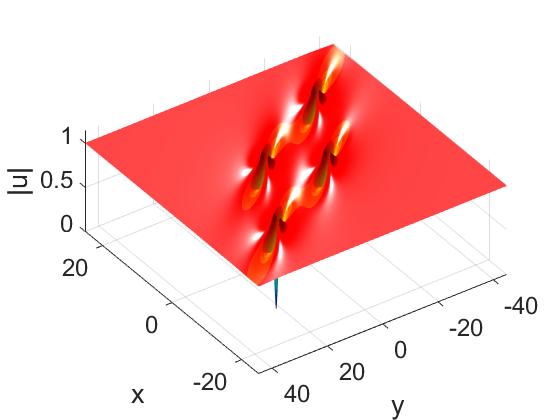}}\newline
\subfigure[]{\includegraphics[height=2.5cm,width=4cm]{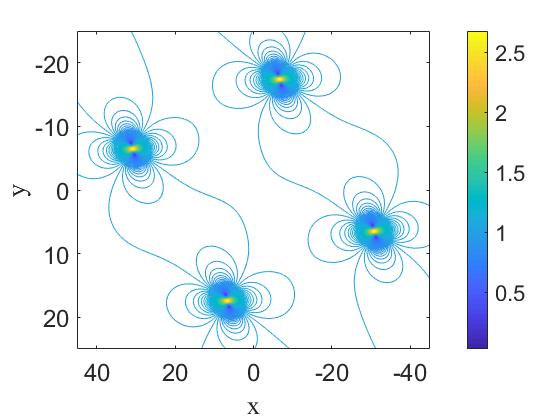}} %
\subfigure[]{\includegraphics[height=2.5cm,width=4cm]{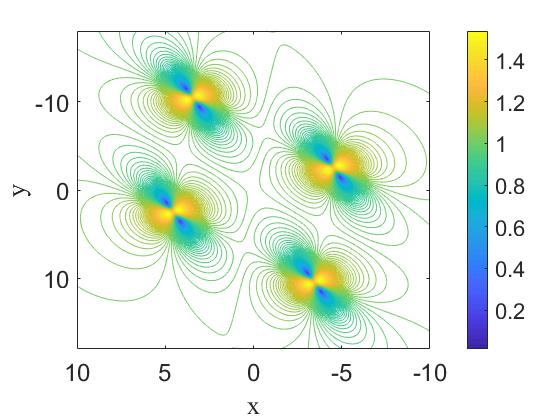}} %
\subfigure[]{\includegraphics[height=2.5cm,width=4cm]{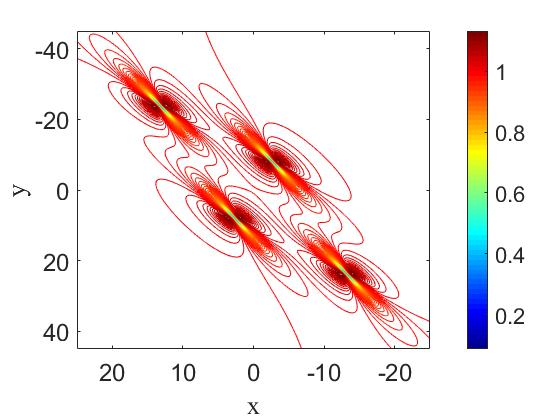}}
\caption{Three species of four-lump solutions of the nonlocal Maccari
system, displayed at $t=0$: (a) a four-bright-lump solution for parameters $%
p_{1}=1+\frac{1}{3}i$, $p_{2}=-1+\frac{1}{4}i$, $\protect\lambda =1$, and $%
s=-1$; (b) a four--lump-four-petal solution for parameters $p_{1}=1+i$, $%
p_{2}=-1+\frac{3}{4}i$, $\protect\lambda =1$, and $s=-1$; (c)\ a
four-dark-lump solution for parameters $p_{1}=\frac{1}{2}+i$, $p_{2}=-\frac{1%
}{2}+\frac{7}{8}i$, $\protect\lambda =1$, and $s=-1$.}
\label{4lump}
\end{figure}

\subsection{General lump-soliton solutions}
\begin{figure}[tbh]
\centering
\subfigure[t=-7]{\includegraphics[height=2.5cm,width=3.5cm]{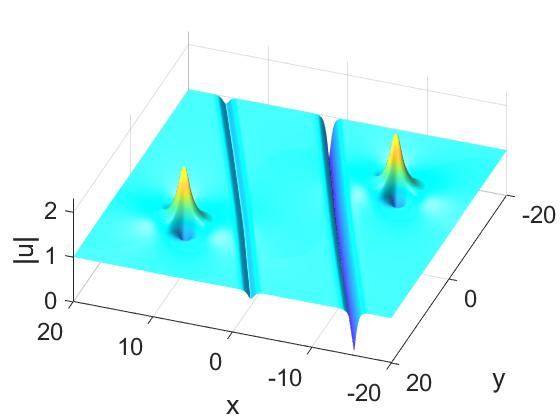}} %
\subfigure[t=-1]{\includegraphics[height=2.5cm,width=3.5cm]{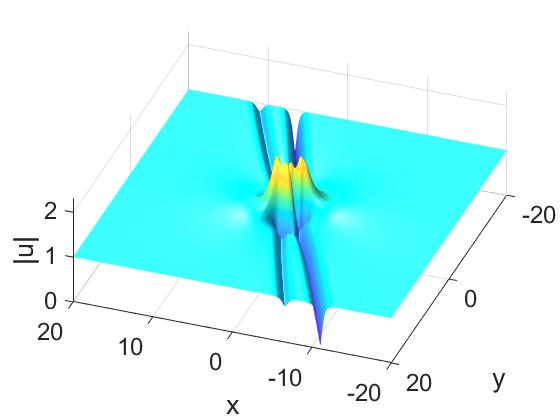}} %
\subfigure[t=0]{\includegraphics[height=2.5cm,width=3.5cm]{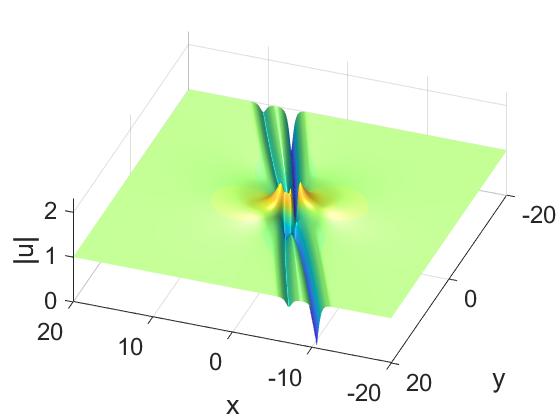}} %
\subfigure[t=9]{\includegraphics[height=2.5cm,width=3.5cm]{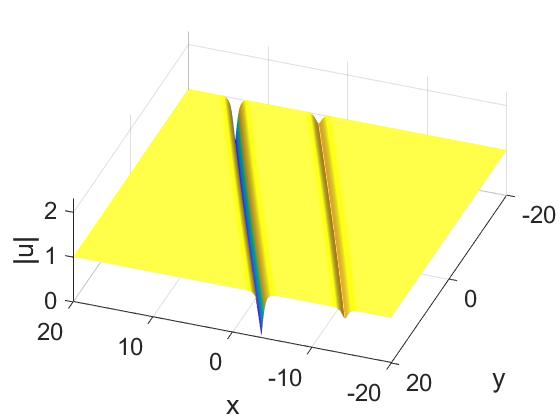}}
\caption{The evolution of a semi-rational solution $|u|$ with parameters
given by eq. \eqref{canshu1}, exhibiting fusion of two lumps and two dark
line solitons into two dark line ones.}
\label{2l2s-1}
\end{figure}

\begin{figure}[tbh]
\centering
\subfigure[t=-9]{\includegraphics[height=2.5cm,width=3.5cm]{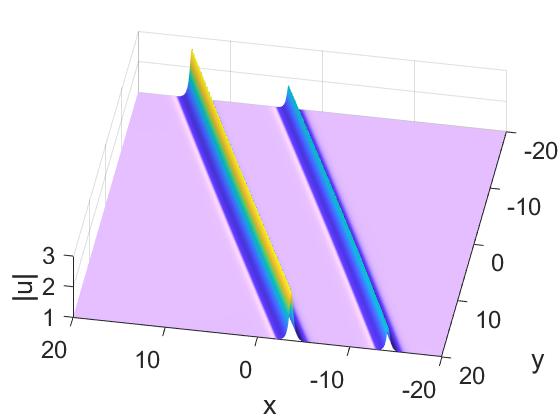}} %
\subfigure[t=-1]{\includegraphics[height=2.5cm,width=3.5cm]{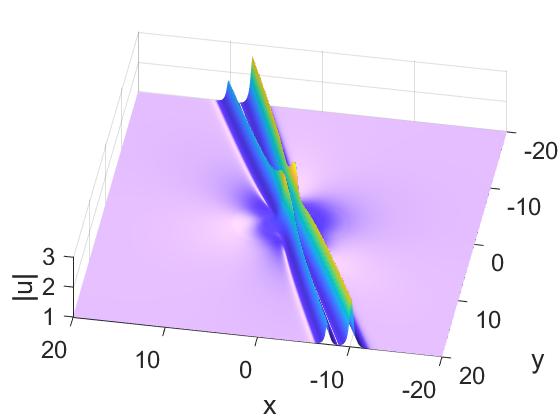}} %
\subfigure[t=1]{\includegraphics[height=2.5cm,width=3.5cm]{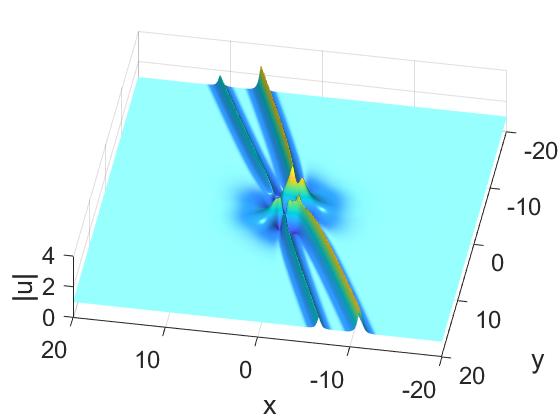}} %
\subfigure[t=9]{\includegraphics[height=2.5cm,width=3.5cm]{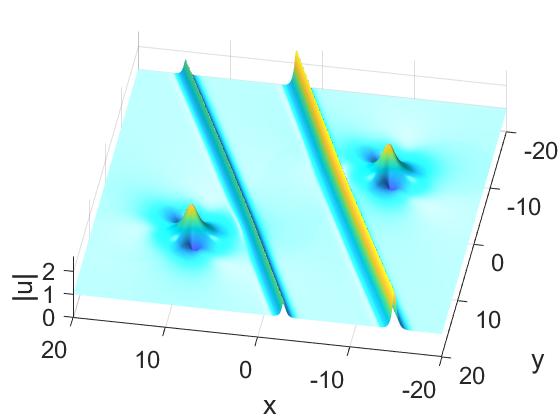}}
\caption{The evolution of a semi-rational solution $|u|$ with parameters
given by eq. \eqref{canshu2}, describing the fission of two anti-dark line
solitons into two lumps and two anti-dark line solitons.}
\label{2l2s-2}
\end{figure}

To illustrated the dynamical behavior of semi-rational solutions composed of
$2N$-lumps and $2N$-solitons, we first consider the case of $N=1,n_{0}=1$ in
eq. \eqref{theor}, with
\begin{equation}\label{2lump-2soliton}
\begin{aligned}
f&=\begin{vmatrix}
\frac{e^{\xi_1+\eta_1}}{p_1+p_1^{*}}[\zeta_1\zeta_1^{'}+\frac{p_1p_1^{*}}{(p_1+p_1^{*})^2}]+\widetilde{c}_{11}|c_{11}|^2 & \frac{e^{\xi_1+\eta_2}}{p_1+p_2^{*}}[\zeta_1\zeta_2^{'}+\frac{p_1p_2^{*}}{(p_1+p_2^{*})^2}]+ib_{12}|c_{11}|^2
\\ \frac{e^{\xi_2+\eta_1}}{p_2+p_1^{*}}[\zeta_2\zeta_1^{'}+\frac{p_2p_1^{*}}{(p_2+p_1^*)^2}]+ib_{21}|c_{11}|^2 & \frac{e^{\xi_2+\eta_2}}{p_2+p_2^{*}}[\zeta_2\zeta_2^{'}+\frac{p_2p_2^{*}}{(p_2+p_2^{*})^2}]+\widetilde{c}_{22}|c_{11}|^2 \end{vmatrix},\\
g&=\begin{vmatrix}
\frac{e^{\xi_1+\eta_1}}{p_1+p_1^{*}}[(\zeta_1+1)(\zeta_1^{'}-1)+\frac{p_1p_1^{*}}{(p_1+p_1^{*})^2}]+\widetilde{c}_{11}|c_{11}|^2 & \frac{e^{\xi_1+\eta_2}}{p_1+p_2^{*}}[(\zeta_1+1)(\zeta_2^{'}-1)+\frac{p_1p_2^{*}}{(p_1+p_2^{*})^2}]+ib_{12}|c_{11}|^2
 \\ \frac{e^{\xi_2+\eta_1}}{p_2+p_1^{*}}[(\zeta_2+1)(\zeta_1^{'}-1)+\frac{p_2p_1^{*}}{(p_2+p_1^{*})^2}]+ib_{21}|c_{11}|^2 & \frac{e^{\xi_2+\eta_2}}{p_2+p_2^{*}}[(\zeta_2+1)(\zeta_2^{'}-1)+\frac{p_2p_2^{*}}{(p_2+p_2^{*})^2}]+\widetilde{c}_{22}|c_{11}|^2
 \end{vmatrix},\\
 \end{aligned}
\end{equation}
where $\zeta _{s}$,$\eta _{j}$,$\zeta ^{_{s}^{{\prime }}}$,$\eta ^{_{j}^{{%
\prime }}}$ $(s,j=1,2)$ are defined in eqs. \eqref{para} and \eqref{zeta},
where $p_{2}=-p_{1}$, $\widetilde{c}_{22}=-\widetilde{c}_{11}^{\ast }$, $%
b_{12}=b_{21}$, $\widetilde{\xi }_{2}=-\widetilde{\xi }_{1}$, $p_{1}$, $%
c_{11}$, $\widetilde{c}_{11}$, $\widetilde{\xi }_{1}$ being complex
parameters, and $b_{12}$ is a real one. Further, setting
\begin{equation}
p_{1}=1+\frac{i}{2},p_{2}=-1-\frac{i}{2},\widetilde{c}_{11}=1+i,\widetilde{c}%
_{22}=-1+i,b_{1,2}=b_{2,1}=0,c_{11}=1,\lambda =1,s=-2,  \label{canshu1}
\end{equation}%
in eq. \eqref{2lump-2soliton}, it is seen in Fig. \eqref{2l2s-1} that the
corresponding semi-rational solution $|u|$ describes fusion of two dark
solitons and two bright lumps into two dark solitons. At the initial stage
of the evolution, two bright lumps are observed on top of the background of
two dark solitons. Interaction between lumps and solitons begins in the
course of the evolution. Eventually, two -bright lumps merge into two-dark
solitons at $t\rightarrow +\infty $. This dynamical phenomenology has not
been previously reported in works dealing with other nonlocal systems.
When taking
\begin{equation}
p_{1}=1+\frac{i}{2},p_{2}=-1-\frac{i}{2},\widetilde{c}_{11}=1+i,\widetilde{c}%
_{22}=-1+i,b_{1,2}=b_{2,1}=0,c_{11}=1,\lambda =1,s=-2,  \label{canshu2}
\end{equation}%
in eq. \eqref{2lump-2soliton}, it is shown in Fig. \eqref{2l2s-2} that the
corresponding semi-rational solution $|u|$ describe fission of two anti-dark
solitons into two bright lumps and two anti-dark solitons. At the initial
stage, the solution features two anti-dark solitons on top of the constant
background. In the course of the evolution, two bright lumps arise from the
two anti-dark solitons. Eventually, two lumps and two anti-dark solitons are
completely separated.
\section{Discussion and conclusion}\label{7}
In this paper, the nonlocal Maccari system \eqref{nonlocal} is
introduced, featuring the specific $\mathcal{PT}$ symmetry. When $\lambda =0$%
, a new nonlocal DS-type equation \eqref{ds-type} is derived. For $s=0$, the
nonlocal Maccari system \eqref{nonlocal} reduces to the defocusing nonlocal
NLS equation \eqref{NLS}, which may be regarded as a two-dimensional
generalization of the nonlocal NLS equation (\ref{nNLS}). Various exact
solutions are presented for the nonlocal Maccari system, produced by means
of the long-wave limit and KP-hierarchy approach, combined with the Hirota
bilinear method. We have derived solutions for the breather, and breather on
top of the periodic background solutions, applying the bilinear form to the
nonlocal Maccari system. We have also derived hyperbolic line-RW
(rogue-wave) solutions, and hyperbolic line-RW ones on top of the background
of periodic line waves, considering the long-wave limit. Nontrivial dynamics
exhibited by these solutions is displayed in Figs. \ref{1b}-\ref{2r1p}.

Furthermore, by constructing different tau-functions of the KP hierarch, general line-soliton solutions with elastic
collisions between the solitons
are derived, as shown in Figs. \ref{2soliton} and \ref{4soliton}.
Additionally, the semi-rational solutions, which consist of $2N$ lumps and $%
2N$ solitons, are presented in terms of the KP-hierarchy method. Their novel
dynamic behavior is shown in Figs. \ref{2l2s-1} and \ref{2l2s-2} These
lump-soliton solutions reduce to $2N$-lump solutions of the nonlocal Maccari
system, for appropriate parameters.

Main differences between the exact solutions of the nonlocal Maccari system
and other nonlocal equations, such as the nonlocal DS (Davey-Stewartson)
equation, nonlocal KP-type system, and nonlocal $(2+1)$-dimensional NLS
equation, are summarized as follows:

\begin{itemize}
\item Periodic solutions. The nonlocal Maccari system \eqref{nonlocal}
produces breathers, and breathers on top of the background of periodic line
waves. The nonlocal DS-I equation gives rise to periodic solitons \cite{ds6}
and line breather solutions \cite{s6,ds3}. The nonlocal (2+1)-dimensional
NLS equation produces line breather solution\cite{ca}.

\item Soliton solutions. The nonlocal Maccari system \eqref{nonlocal}
generates three kinds of two-soliton solutions, \textit{viz}., dark-dark
solitons, dark anti-dark solitons, and anti-dark anti-dark solitons. In
these complexes, two bound solitons stay parallel to each other. This
features is not found in the nonlocal DS equation and (2+1)-dimensional NLS
equation. The nonlocal DS-I equation produces a dark cross soliton \cite{ds4}
and a double-peak dromion \cite{ds6}.

\item Rational solutions. The second-order rational solution of nonlocal Maccari
system\eqref{nonlocal} is a hyperbolic line RW and 2-lump solutions.
However, the second-order rational solution of the nonlocal DS-II equation
features a cross-shape line RW \cite{ds5} and dark-anti-dark rational
travelling waves \cite{aml}.

\item Semi-rational solutions. We have presented: (a) hyperbolic line RWs on
top of the background of periodic line waves; (b) fusion of two line
solitons and lumps into line solitons; (c) fission of two line solitons into
lumps and line solitons of the nonlocal Maccari system \eqref{nonlocal}. The
semi-rational solutions of the nonlocal 2D NLS equation are composed of a
cross-shaped line RW and periodic line waves \cite{mu3}. The semi-rational
solutions of the nonlocal KP-type system are composed of a soliton and two
lumps \cite{ks3}.
\end{itemize}

These new results enrich the structure of waves in nonlocal systems, which
suggest that the nonlocal Maccari system \eqref{nonlocal} is quite a
relevant nonlocal extension of the nonlocal NLS equation (\ref{nNLS}). The
system also provides a useful model for understanding new features of
nonlinear dynamics of $\mathcal{PT}$-symmetric systems.
\section*{Acknowledgments}
This work is supported by the NSF of China under Grants No. 11671219, No.
11871446, No.12071304 and No.12071451.
The work of B.A.M. is supported, in part, by the Israel Science
Foundation through grant No. 1286/17.

\appendix
	\section{} In this appendix, we give the proof of Lemma 2.
	
\textbf{Proof} As
\begin{equation}
\begin{aligned}
(\xi_j+\eta_j)(x,y,t)&=(p_j+p_j^*+\frac{1}{p_j}+\frac{1}{p^*_j})x+
\left(\frac{\lambda(p_j^2+1)\sqrt{-s}-p_j^2+1}{2\sqrt{-s}p_j}
+\frac{\lambda(p_j^{*2}+1)\sqrt{-s}-p_j^{*2}+1}{2\sqrt{-s}p_j^*}\right)y\\
&+\frac{i}{2}(p_j^2-\frac{1}{p_j^2}+p_j^{*2}-\frac{1}{p_j^{*2}})t+\xi_{j0}+\eta_{j0},\\
\end{aligned}
\end{equation}
and
\begin{equation}
\begin{aligned}
(\xi_{M+j}+\eta_{M+j})(x,y,t)=&(p_{M+j}+p_{M+j}^*+\frac{1}{p_{M+j}}+\frac{1}{p^*_{M+j}})x+
(\frac{\lambda(p_{M+j}^2+1)\sqrt{-s}-p_{M+j}^2+1}{2\sqrt{-s}p_{M+j}}\\
&+\frac{\lambda(p_{M+j}^{*2}+1)\sqrt{-s}-p_{M+j}^{*2}+1}{2\sqrt{-s}p_{M+j}^*})y
+\frac{i}{2}(p_{M+j}^2-\frac{1}{p_{M+j}^2}+p_{M+j}^{*2}-\frac{1}{p_{M+j}^{*2}})t\\
&+\xi_{{M+j}0}+\eta_{{M+j}0},\\
=&-(p_j+p_j^*+\frac{1}{p_j}+\frac{1}{p^*_j})x-
\left(\frac{\lambda(p_j^2+1)\sqrt{-s}-p_j^2+1}{2\sqrt{-s}p_j}
+\frac{\lambda(p_j^{*2}+1)\sqrt{-s}-p_j^{*2}+1}{2\sqrt{-s}p_j^*}\right)y\\
&+\frac{i}{2}(p_j^2-\frac{1}{p_j^2}+p_j^{*2}-\frac{1}{p_j^{*2}})t+\xi_{j0}+\eta_{j0},
\end{aligned}
\end{equation}
so we can obtain
\begin{equation}
\begin{aligned}
(\xi_{M+j}+\eta_{M+j})^*(-x,-y,t)=(\xi_j+\eta_j)(x,y,t),\\
(\xi_j+\eta_j)^*(-x,-y,t)=(\xi_{M+j}+\eta_{M+j})(x,y,t).
\end{aligned}
\end{equation}
Besides, the tau function can be rewritten as
\begin{align*}
\tau_{n}(x,y,t)=\prod_{k=1}^{2M}e^{\xi_k+\xi^*_k} &\begin{vmatrix} c_j\delta_{jk}e^{-(\xi_j+\xi^*_k)}
+(-\frac{p_j}{p^*_k})^n\frac{1}{p_{j}+p^*_{k}} & (-\frac{p_j}{p^*_{M+k}})^n\frac{1}{p_{j}+p^*_{M+k}}
 \\ (-\frac{p_{M+j}}{p^*_k})^n\frac{1}{p_{M+j}+p^*_{k}} & c_{M+j}\delta_{M+j,M+k}e^{-(\xi_{M+j}+\xi^*_{M+k})}
+(-\frac{p_{M+j}}{p^*_{M+k}})^n\frac{1}{p_{M+j}+p^*_{M+k}} \end{vmatrix} \\
=\prod_{k=1}^{2M}e^{\xi_k+\xi^*_k} &\begin{vmatrix} c_{M+j}\delta_{M+j,M+k}e^{-(\xi_{M+j}+\xi^*_{M+k})}
+(-\frac{p_{M+j}}{p^*_{M+k}})^n\frac{1}{p_{M+j}+p^*_{M+k}} & (-\frac{p_{M+j}}{p^*_k})^n\frac{1}{p_{M+j}+p^*_{k}}
 \\ (-\frac{p_j}{p^*_{M+k}})^n\frac{1}{p_{j}+p^*_{M+k}} & c_j\delta_{jk}e^{-(\xi_j+\xi^*_k)}
+(-\frac{p_j}{p^*_k})^n\frac{1}{p_{j}+p^*_{k}} \end{vmatrix}\\
=\prod_{k=1}^{2M}e^{\xi_k+\xi^*_k} &\begin{vmatrix} c^*_{j}\delta_{M+j,M+k}e^{-(\xi_{M+j}+\xi^*_{M+k})}
+(-\frac{p_j}{p^*_k})^n\frac{1}{p_{j}+p^*_{k}} & (-\frac{p_j}{p^*_{M+k}})^n\frac{1}{p_{j}+p^*_{M+k}}
\\ (-\frac{p_{M+j}}{p^*_k})^n\frac{1}{p_{M+j}+p^*_{k}} & c^*_{M+j}\delta_{jk}e^{-(\xi_j+\xi^*_k)}
+(-\frac{p_{M+j}}{p^*_{M+k}})^n\frac{1}{p_{M+j}+p^*_{M+k}} \end{vmatrix},
\end{align*}
On the other hand
\begin{align*}
\tau_{n}(x,y,t)=\prod_{k=1}^{2M}e^{\xi_k+\xi^*_k} &\begin{vmatrix} c_j\delta_{jk}e^{-(\xi_j+\xi^*_k)}
+(-\frac{p_j}{p^*_k})^n\frac{1}{p_{j}+p^*_{k}} & (-\frac{p_j}{p^*_{M+k}})^n\frac{1}{p_{j}+p^*_{M+k}}
 \\ (-\frac{p_{M+j}}{p^*_k})^n\frac{1}{p_{M+j}+p^*_{k}} & c_{M+j}\delta_{M+j,M+k}e^{-(\xi_{M+j}+\xi^*_{M+k})}
+(-\frac{p_{M+j}}{p^*_{M+k}})^n\frac{1}{p_{M+j}+p^*_{M+k}} \end{vmatrix} \\
=\prod_{k=1}^{2M}e^{\xi_k+\xi^*_k} &\begin{vmatrix} c_k\delta_{kj}e^{-(\xi_k+\xi^*_j)}
+(-\frac{p_k}{p^*_j})^n\frac{1}{p_{k}+p^*_{j}} & (-\frac{p_k}{p^*_{M+j}})^n\frac{1}{p_{k}+p^*_{M+j}}
 \\ (-\frac{p_{M+k}}{p^*_j})^n\frac{1}{p_{M+k}+p^*_{j}} & c_{M+k}\delta_{M+k,M+j}e^{-(\xi_{M+k}+\xi^*_{M+j})}
+(-\frac{p_{M+k}}{p^*_{M+j}})^n\frac{1}{p_{M+k}+p^*_{M+j}} \end{vmatrix} ,
\end{align*}
thus
\begin{equation}
\begin{aligned}
\tau^*_{n}=&\prod_{k=1}^{2M}e^{\theta_1}
\begin{vmatrix} c^*_k\delta_{kj}e^{-(\xi_k+\xi^*_j)^*(-x,-y,t)}
+(-\frac{p^*_k}{p_j})^n\frac{1}{p^*_{k}+p_{j}} & (-\frac{p^*_k}{p_{M+j}})^n\frac{1}{p^*_{k}+p_{M+j}}
 \\ (-\frac{p^*_{M+k}}{p_j})^n\frac{1}{p^*_{M+k}+p_{j}} &
 c^*_{M+k}\delta_{M+k,M+j}e^{-\theta_2}
+(-\frac{p^*_{M+k}}{p_{M+j}})^n\frac{1}{p^*_{M+k}+p_{M+j}} \end{vmatrix}\\
=&\prod_{k=1}^{2M}e^{(\xi_k+\xi^*_k)}
\begin{vmatrix} c^*_k\delta_{kj}e^{-(\xi_{M+k}+\xi^*_{M+j})}
+(-\frac{p^*_k}{p_j})^n\frac{1}{p^*_{k}+p_{j}} & (-\frac{p^*_k}{p_{M+j}})^n\frac{1}{p^*_{k}+p_{M+j}}
 \\ (-\frac{p^*_{M+k}}{p_j})^n\frac{1}{p^*_{M+k}+p_{j}} &
 c^*_{M+k}\delta_{M+k,M+j}e^{-(\xi_{k}+\xi^*_{j}))}
+(-\frac{p^*_{M+k}}{p_{M+j}})^n\frac{1}{p^*_{M+k}+p_{M+j}} \end{vmatrix},
\end{aligned}
\end{equation}
where $\tau^*_{n}=\tau^*_{n}(-x,-y,t), \theta_1=(\xi_k+\xi^*_k)^*(-x,-y,t),
\theta_2=(\xi_{M+k}+\xi^*_{M+j})^*(-x,-y,t)$. Obviously $\tau^*_{n}(-x,-y,t)=\tau_{n}(x,y,t)$.

\section{} In this appendix, we give the proof of Lemma 3.

\textbf{Proof} Since $p_{N+j}=-p_j$,$q_j=p^*_j$,
$\widetilde{\eta}_{j}=-\widetilde{\xi}^*_{j}$,
$\widetilde{\xi}_{N+s}=\widetilde{\xi}_{s}$,
we have
\begin{equation}
\begin{aligned}
&(\xi_{N+s}+\eta_{N+j})^*(-x,-y,t)=(\xi_j+\eta_s)(x,y,t),\\
&(\xi_{s}+\eta_{N+j})^*(-x,-y,t)=(\xi_{N+j}+\eta_s)(x,y,t),\\
&\xi^{'*}_{N+s}(-x,-y,t)=\eta^{'}_s(x,y,t),\\
&\eta^{'*}_{N+s}(-x,-y,t)=\xi^{'}_s(x,y,t).
\end{aligned}
\end{equation}
Furthermore
\begin{equation}
\begin{aligned}
M_{N+s,N+j}^{(n)*}(-x,-y,t)&=(-\frac{p^*_{N+s}}{q^*_{N+j}})^ne^{(\xi_{N+s}
+\eta_{N+j})^*(-x,-y,t)}\sum_{k=0}^{n_0}c^*_{N+s,N+k}
(p^*_{N+s}\partial_{p^*_{N+s}}+\xi^{'*}_{N+s}(-x,-y,t)+n)^{n_0-k}\\
& \times \sum_{l=0}^{n_0}d^*_{N+j,N+l}(q^*_{N+j}
\partial_{q^*_{N+j}}+\eta^{'*}_{N+j}(-x,-y,t)-n)^{n_0-l}\frac{1}{p^*_{N+s}+q^*_{N+j}}
+\widetilde{c}^*_{N+s,N+j}\\
&=-(-\frac{p_j}{q_s})^{-n}e^{\xi_j+\eta_s}\sum_{k=0}^{n_0}d_{sk}
(q_s\partial_{q_s}+\eta^{'}_s+n)^{n_0-k}\\
& \times \sum_{l=0}^{n_0}c_{jl}(p_j
\partial_{p_j}+\xi^{'}_j-n)^{n_0-l}\frac{1}{p_j+q_s}-\widetilde{c}_{js}\\
&=-M_{j,s}^{(-n)}(x,y,t),
\end{aligned}
\end{equation}
similarly
\begin{equation}
\begin{aligned}
M_{s,j}^{(n)*}(-x,-y,t)=-M_{N+j,N+s}^{(-n)}(x,y,t).
\end{aligned}
\end{equation}
On the other hand
\begin{equation}
\begin{aligned}
M_{s,N+j}^{(n)*}(-x,-y,t)&=(-\frac{p^*_{s}}{q^*_{N+j}})^ne^{(\xi_{s}
+\eta_{N+j})^*(-x,-y,t)}\sum_{k=0}^{n_0}c^*_{s,N+k}
(p^*_{s}\partial_{p^*_{s}}+\xi^{'*}_{s}(-x,-y,t)+n)^{n_0-k}\\
& \times \sum_{l=0}^{n_0}d^*_{N+j,l}(q^*_{N+j}
\partial_{q^*_{N+j}}+\eta^{'*}_{N+j}(-x,-y,t)-n)^{n_0-l}\frac{1}{p^*_{s}+q^*_{N+j}}
+\widetilde{c}^*_{s,N+j}\\
&=-(-\frac{p_j}{q_{N+s}})^{-n}e^{\xi_j+\eta_{N+s}}\sum_{k=0}^{n_0}d_{N+s,k}
(q_{N+s}\partial_{q_{N+s}}+\eta^{'}_{N+s}+n)^{n_0-k}\\
& \times \sum_{l=0}^{n_0}c_{jl}(p_j
\partial_{p_j}+\xi^{'}_j-n)^{n_0-l}\frac{1}{p_j+q_{N+s}}-\widetilde{c}_{j,N+s}\\
&=-M_{j,N+s}^{(-n)}(x,y,t),
\end{aligned}
\end{equation}
similarly
\begin{equation}
\begin{aligned}
M_{N+s,j}^{(n)*}(-x,-y,t)=-M_{N+j,s}^{(-n)}(x,y,t).
\end{aligned}
\end{equation}
These results imply
\begin{equation}
\begin{aligned}
\tau^*_n(-x,-y,t)&=\begin{vmatrix} M_{s,j}^{(n)*}(-x,-y,t) & M_{s,N+j}^{(n)*}(-x,-y,t)
 \\ M_{N+s,j}^{(n)*}(-x,-y,t) & M_{N+s,N+j}^{(n)*}(-x,-y,t) \end{vmatrix}\\
&=\begin{vmatrix} -M_{N+s,N+j}^{(-n)}& -M_{j,N+s}^{(-n)}
 \\ -M_{N+j,s}^{(-n)} & -M_{j,s}^{(-n)}\end{vmatrix}\\
&=\begin{vmatrix} M_{j,s}^{(-n)}& M_{N+j,s}^{(-n)}
 \\ M_{j,N+s}^{(-n)} & M_{N+j,N+s}^{(-n)}\end{vmatrix}\\
&=\tau_{-n}(x,y,t),
 \end{aligned}
\end{equation}
this completes the proof.

\begin{thebibliography}{99}
\bibitem{ablow}M. J. Ablowitz, P. A. Clarkson, Nonlinear evolution equations and inverse scattering, Cambridge University Press, 1991.

\bibitem{hirota}R. Hirota, The direct method in soliton theory, Cambridge University Press, 2004.

\bibitem{waz}A. M. Wazwaz, Partial differential equations and solitary waves theory (nonlinear physical science), Berlin: Springer, 2009.

\bibitem{r1}Y. Ohta, J. K. Yang, Proc. R. Soc. A 468 (2012) 1716-1740.

\bibitem{r2}P. Dubard, P. Gaillard, C. Klein, V. B. Matveev, Eur. Phys. J. Spec. Top. 185 (2010) 247-258.

\bibitem{r3}N. Akhmediev, A. Ankiewicz, J. M. Soto-Crespo, Phys. Rev. E 80 (2009) 026601.

\bibitem{r4} J. S. He, H. R. Zhang, L. H. Wang, K. Porsezian, A. S. Fokas, Phys. Rev. E 87 (2013) 052914.

\bibitem{r5}J. S. He, S. W. Xu, K. Porsezian, Phys. Rev. E 86 (2012) 066603.

\bibitem{r6}J. C. Chen, Y. Chen, B. F. Feng, Phys. Lett. A 379 (2015) 1510-1519.

\bibitem{r7}D. Mihalache, Rom. Rep. Phys. 69 (2017) 403.

\bibitem{r8}S. H. Chen, F. Baronio, J. M. Soto-Crespo, P. Grelu, D. Mihalache, J. Phys. A: Math. Theor. 50 (2017) 463001.

\bibitem{r9}L. M. Ling, L. C. Zhao, Z. Y. Yang, B. L. Guo, Phys. Rev. E 96 (2017) 022211.

\bibitem{nrw1}Z. Y. Yan, Phys. Lett. A 374 (2010) 672-679.

\bibitem{nrw2}X. Y. Wen, Z. Y. Yan, Commun. Nonlinear Sci. Numer. Simulat. 43 (2017) 311-329.

\bibitem{pe}D. H. Peregrine, J. Aust. Math. Soc. B 25 (1983) 16-43.

\bibitem{guolj}L. J. Guo, L. H. Wang, Y. Cheng, J. S. He, Commun. Nonlinear Sci. Numer. Simul. 52 (2017) 11-23.

\bibitem{zhangys}Y. S. Zhang, D. Q. Qiu, D. Mihalache, J. S. He, Chaos 28 (2018) 103108.

\bibitem{fengbf}B. F. Feng, L. M. Ling, D. A. Takahashi, Stud. Appl. Math. 144, (2019) 46-101.

\bibitem{ma1}B. Ren, W. X. Ma, J. Yu, Nonlinear Dyn. 96 (2019) 717-727.

\bibitem{ma2}W. X. Ma, J. Li, C. M. Khalique, Complexity (2018) 9059858.

\bibitem{be}Y. Bludov, V. Konotop, N. Akhmediev, Phys. Rev. A 80 (2009) 033610.

\bibitem{os1}J. Dudley, F. Dias, M. Erkintalo, G. Genty, Nature Photonics 8 (2014) 755-764.

\bibitem{os2}A. Montina, U. Bortolozzo, S. Residori, F. Arecchi, Phys. Rev. Lett. 103 (2009) 173901.

\bibitem{super}A. N. Ganshin, V. B. Efimov, G. V. Kolmakov, L. P. Mezhov-Deglin, P. V. E. McClintock, Phys. Rev. Lett. 101 (2008) 065303.

\bibitem{fi}Z. Y. Yan, Phys. Lett. A 375 (2011) 4274-4279.

\bibitem{s1}G. Mu, Z. Y. Qin, Nonlinear Anal. RWA 31 (2016) 179.

\bibitem{s2}A. Degasperis, S. Lombardo, Phys. Rev. E 88 (2013) 052914.

\bibitem{s3}B. L. Guo, L. M. Ling, Chinese Phys. Lett. 28 (2011) 110202.

\bibitem{s4}F. Baronio, A. Degasperis, M. Conforti, S. Wabnitz, Phys. Rev. Lett. 109 (2012) 044102.

\bibitem{s5}J. G. Rao, K. Porsezian, J. S. He, Chaos 27 (2017) 083115.

\bibitem{s6}J. G. Rao, Y. S. Zhang, A. S. Fokas, J. S. He, Nonlinearity 31 (2018)
  4090-4107.

\bibitem{s7}J. B. Chen, D. E. Pelinovsky, R. E. White, Phys. Rev. E 100 (2019) 052219.

\bibitem{semi1}Z. Y. Yan, V. V. Konotop, N. Akhmediev, Phys. Rev. E 82 (2010) 036610.

\bibitem{semi2}G. Q. Zhang, Z. Y. Yan, L. Wang, Proc. R. Soc. A 475 (2019) 20180625.

\bibitem{nls}M. J. Ablowitz, Z. H. Musslimani, Phys. Rev. Lett. 110 (2013) 064105.

\bibitem{dispt1}P. G. Kevrekidis, D. E. Pelinovsky, D. Y. Tyugin, SIAM J. Appl. Dyn. Syst. 12 (2013) 1210-1236.

\bibitem{dispt2}P. G. Kevrekidis, D. E. Pelinovsky, D. Y. Tyugin, J. Phys. A: Math. Theor. 46 (2013) 365201.

\bibitem{dispt3}D. E. Pelinovsky, D. A. Zezyulin, V. V. Konotop, J. Phys. A: Math. Theor. 47 (2014) 085204.

\bibitem{dispt4}A. Chernyavsky, D. E. Pelinovsky, Physica D 371 (2018) 48-59.

\bibitem{i1}X. Huang, L. M. Ling, Eur. Phys. J. Plus 131 (2016) 148.

\bibitem{i2}X. Y. Wen, Z. Yan, Y. Yang, Chaos 26 (2016) 063123.

\bibitem{i3}V. S. Gerdjikov, A. Saxena, J. Math. Phys. 58 (2017) 013502.

\bibitem{i4}A. K. Sarma, M. A. Miri, Z. H. Musslimani, D. N. Christodoulides, Phys. Rev. E 89 (2014) 052918.

\bibitem{i5}M. Li, T. Xu, Phys. Rev. E 91 (2015) 033202.

\bibitem{i6}L. Y. Ma, Z. N. Zhu, J. Math. Phys. 57 (2016) 083507.

\bibitem{mu1}A. S. Fokas, Nonlinearity 29 (2016) 319-324.

\bibitem{mu2}M. J. Ablowitz, Z. H. Musslimani, Stud. Appl. Math. 139 (2017) 7-59.

\bibitem{mu3}Y. L. Cao, J. G. Rao, D. Mihalache, J. S. He, Appl. Math. Lett. 80 (2018) 27-34.

\bibitem{mu4}W. Liu, X. X. Zhang, X. L. Li, Nonlinear Dyn. 94 (2018) 2177-2189.

\bibitem{mu5}Y. Shi, Y. S. Zhang, S. W. Xu, Nonlinear Dyn. 94 (2018) 2327-2334.

\bibitem{mu6}Z. Y. Yan, Appl. Math. Lett. 47 (2015) 61-68.

\bibitem{mu7}L. Y. Ma, S. F. Tian, Z. N. Zhu, J. Math. Phys. 58 (2017) 103501.

\bibitem{WE}W. Eckhaus, Lecture Notes in Physics 246 (1986) 168-194.

\bibitem{FC}F. Calogero, W. Eckhaus, Inverse Prob. 3 (1987) 27-32.

\bibitem{FC1}F. Calogero, W. Eckhaus, Inverse Prob. 3 (1987) 229--262.

\bibitem{FC2}F. Calogero, W. Eckhaus, Inverse Prob. 4 (1988) 11-33.

\bibitem{mac}A. Maccari, J. Math. Phys. 37 (1996) 6207-6212.

\bibitem{maccari}A. Maccari, Phys. Lett. A 384 (2020) 126740.

\bibitem{ks1}H. L. Zhen, B. Tian, W. R. Song, Appl. Math. Lett. 27 (2014) 90-96.

\bibitem{ks2}Y. S. Zhang, J. G. Rao, K. Porsezian, J. S. He, Nonlinear Dyn 95 (2019) 1133-1146.

\bibitem{ks3}J. G. Rao, J. S. He, D. Mihalache, Y. Cheng, Appl. Math. Lett. 94 (2019) 166-173.

\bibitem{ks4}G. H. Wang, L. H. Wang, J. G. Rao, J. S. He, Commu. Theor. Phys. 67 (2017) 601-610.

\bibitem{ks5}W. Liu, A. M. Wazwaz, Rom. J. Phys. 64 (2019) 111.

\bibitem{ks6}A. M. Wazwaz, Phys. Scri. 85 (2012) 065011.

\bibitem{sh}E. I. Shulman, Theor. Math. Phys. 56 (1983) 720.

\bibitem{dstype}A. Davey, K. Stewartson, Proc. R. Soc. London A 338 (1974) 101-110.

\bibitem{dstype1}D. Anker, N. C. Freeman, Proc. R. Soc. London A 360 (1978) 529-540.

\bibitem{ds1}C. Rogers, K. W. Chow, R. Conte, Nuovo Cimento B 122 (2007) 105-111.

\bibitem{ds2}A. S. Fokas, Comm. Math. Phys. 289 (2009) 957-993.

\bibitem{ds3}J. G. Rao, Y. Cheng, J. S. He, Stud. Appl. Math. 139 (2017) 568-598.

\bibitem{ds4}Z. X. Zhou, Stud. Appl. Math. 141 (2018) 186-204.

\bibitem{ds5}B. Yang, Y. Chen, Commun. Nonlinear Sci. Numer. Simul. 69 (2019) 287-303.

\bibitem{ds6}T. Xu, M. Li, Y. Huang, Y. Chen, C. Yu, Modern Phys. Lett. B 31 (2017) 1750388.

\bibitem{ds7}Y. Ohta, J. Yang, Phys. Rev. E 86 (2012) 036604.

\bibitem{ds8}A. Fokas, D. Pelinovsky, C. Sulem, Physica D 152 (2001) 189-198.

\bibitem{lin}A. Issasfa, J. Lin, Inter. J. Mod. Phys. B 33 (2019) 1950317.

\bibitem{bs1}C. M. Bender, D. C. Brody, H. F. Jones, Phys. Rev. Lett. 89 (2002) 270401.

\bibitem{bs2}C. M. Bender, S. Boettcher, Phys. Rev. Lett. 80 (1998) 5243-5246.

\bibitem{sy1}L. Feng, Z. J. Wong, R. M. Ma, Y. Wang, X. Zhang, Science 346 (2014) 972.

\bibitem{sy2}A. Regensburger, C. Bersch, M. A. Miri, G. Onishchukov, D. N. Christodoulides, U. Peschel, Nature 488 (2012) 167.

\bibitem{sy3}V. V. Konotop, J. K. Yang, D. A. Zezyulin, Rev. Mod. Phys. 88 (2016) 035002.

\bibitem{le}B. Bagchi, C. Quesne, Phys. Lett. A 273 (2000) 285-292.

\bibitem{lian}C. M. Bender, D. C. Brody, H. F. Jones, Am. J. Phys. 71 (2003) 1095.

\bibitem{jt1}Z. Ahmed, Phys. Rev. A 64 (2001) 042716.

\bibitem{jt2}M. Znojil, Phys. Lett. A 285 (2001) 7-10.

\bibitem{boris}E. Luz, V. Lutsky, E. Granot, B. A. Malomed, Scientific Reports 9 (2019) 4483.

\bibitem{qian}C. Qian, J. G. Rao, Y. B. Liu, J. S. He, Chinese Phys. Lett. 33 (2016) 110201.

\bibitem{raso1}G. Q. Zhang, Z. Y. Yan, Y. Chen, Appl. Math. Lett. 69 (2017) 113-120.

\bibitem{raso2}G. Q. Zhang, Z. Y. Yan, EPL, 118 (2017) 60004.

\bibitem{ca}Y. L. Cao, B. A. Malomed, J. S. He, Chaos Soliton Fract. 114 (2018)
  99-107.

\bibitem{feng1}B. F. Feng, X. D. Luo, M. J. Ablowitz, Z. H. Musslimani, Nonlinearity 31 (2018) 5385-5409.

\bibitem{sun1}B. N. Sun, Nonlinear Dyn. 92 (2018) 1369-1377.

\bibitem{rao1}J. G. Rao, Y. Cheng, K. Porsezian, D. Mihalache, J. S. He, Physica D 401 (2020) 132180.

\bibitem{sato1}M. Jimbo, T. Miwa, Publ. RIMS Kyoto Univ. 19 (1983) 943-1001.

\bibitem{sato2}E. Date, M. Kashiwara, M. Jimbo, T. Miwa, Transformation groups for soliton equations, in: M. Jimbo, T. Miwa (Eds.), Nonlinear Integrable
  Systems-Classical Theory and Quantum Theory, World Scientific, Singapore.

\bibitem{sato3}Y. Ohta, D. Wang, J. Yang, Stud. Appl. Math. 127 (2011) 345-371.

\bibitem{sato5}Y. Ohta, J. Yang, J. Phys. A: Math. Theor. 46 (2013) 105202.

\bibitem{ctp}J. G. Rao, L. H. Wang, Y. Zhang, J. S. He, Commun. Theor. Phys. 64 (2015) 605-618.

\bibitem{aml}B. Yang, Y. Chen, Appl. Math. Lett. 82 (2018) 43-49.
\end{thebibliography}
\end{document}